\documentclass[a4paper,11pt]{article}
\pdfoutput=1 

\usepackage{jcappub} 

\usepackage[T1]{fontenc} 
\usepackage{amsmath}
\usepackage{cleveref}
\usepackage{footmisc}
\usepackage[utf8]{inputenc}

\newcommand{\roughly}[1]{\mathrel{\raise.3ex\hbox{$#1$\kern-0.85em
\lower1ex\hbox{$\sim$}}}}

\newcommand{\be}{\begin{equation}}
\newcommand{\bee}{\begin{equation}}
\newcommand{\ee}{\end{equation}}
\newcommand{\beea}{\begin{eqnarray}}
\newcommand{\eea}{\end{eqnarray}}
\newcommand{\bea}{\begin{eqnarray}}

\def\pref#1{(\ref{#1})}
\def\exd{{\rm d}}

\def\cR{{\cal R}}

\def\ssB{{\scriptscriptstyle B}}

\def\ssH{{\scriptscriptstyle H}}

\def\ssN{{\scriptscriptstyle N}}

\def\ssR{{\scriptscriptstyle R}}

\title{Constraining Fundamental Physics with the Event Horizon Telescope}

\author[a,b]{Markus Rummel}
\author[a,b]{and C.P.~Burgess}
\affiliation[a]{Physics \& Astronomy, McMaster University, Hamilton, ON, Canada,
L8S 4M1}
\affiliation[b]{Perimeter Institute for Theoretical Physics, Waterloo, Ontario 
N2L 2Y5, Canada }
\emailAdd{rummelm@mcmaster.ca}
\emailAdd{cburgess@perimeterinstitute.ca}

\abstract{We show how Event Horizon Telescope (EHT) observations of the
supermassive object at the center of M87 can constrain deviations from General
Relativity (GR) in a relatively model-independent way. We focus on the class of
theories whose deviations from GR modify black holes into alternative compact
objects whose properties approach those of an ordinary black hole sufficiently
far from the would-be event horizon. We examine this class for two reasons:
($i$) they tend to reproduce black-hole expectations for astrophysical
accretion disks (and so do not undermine the evidence linking black holes to
active galactic nuclei); ($ii$) they lend themselves to a robust
effective-field-theory treatment that expands in powers of $\ell/r$, where
$\ell$ is the fundamental length scale that sets the distance over which
deviations from GR are significant and $r$ is a measure of distance from the
would-be horizon. At leading order the observational impact of these types of
theories arise as modifications to the transmission and reflection coefficients
of modes as they approach the horizon. We show how EHT observations can
constrain this reflection coefficient, assuming only that the deviations from
GR are small enough to be treated perturbatively. Our preliminary analysis
indicates that such reflection coefficients can already be constrained to be
less than of order 10\% (corresponding to $\ell \lesssim 100 \mu m$), and so can rule
out some benchmark cases used when seeking black-hole echoes. The precise
bounds depend on the black hole spin, as well as on detailed properties of the
reflection coefficient (such as its dependence on angular direction).}

\begin{document}

\maketitle
\flushbottom

\section{Introduction}
\label{sec:intro}

New observational information very often drives progress in fundamental
science, and the strikingly new information of our time comes from the recent
imaging of black holes using both gravitational waves~\cite{PhysRevX.9.031040}
and the Event Horizon Telescope (EHT)~\cite{Akiyama:2019cqa}. In principle, the
apparent agreement between these observations and the predictions of General
Relativity (GR) provide constraints on the way new fundamental physics might
deviate from GR in the strong-gravity near-horizon regime, though the reliable
extraction of these constraints remains a relatively new field. 

Throughout most of physics new observations are typically used to constrain
fundamental physics in one of two ways. 
\begin{itemize}
\item {\em Model Building:} The first approach considers each detailed proposal
	for new physics on its own terms, computing its implications on a
		model-by-model basis. This is the approach used, for instance,
		when the implications of specific proposals ({\it
		e.g.}~Brans-Dicke scalars, axion models, {\it etc.}) are
		compared with the predictions of GR. 
\item {\em Effective Field Theories (EFTs):} The second approach is appropriate
	if no new degrees of freedom in the modification actually appear in the
		phenomenon of interest (perhaps they are too heavy or
		short-ranged to be relevant to the scales being measured). In
		this case all new degrees of freedom can be integrated out,
		with new effects parameterized using a low-energy/long-distance
		expansion. In this framework new physics enters only through
		short-distance effects (smaller than some characteristic scale
		$\ell$) and builds local interactions using only known fields,
		organized in powers of derivatives times $\ell$.
\end{itemize}

A strength of the model-building approach (which is the direction to this point
most explored in the literature (see, for example, \cite{Berti:2015itd,
Sathyaprakash:2019yqt} for surveys with references) is that individual models
can be very predictive, with many possible observable deviations from GR
characterized by a small number of model parameters. Its main drawback is
inefficiency; there are a great many models from which to choose and at present
we have no idea which is likely to be the right one. So in principle one must
work through them all, making detailed predictions for each. In practice this
means a few popular models get detailed attention and others -- though perhaps
equally deserving -- do not.

The strengths (and weaknesses) of an EFT approach are complementary to those of
model-by-model calculations. The main strength is the robustness of results:
EFTs capture the dominant way that {\it any} model in a very broad category can
affect observations (for reviews relevant to applications with gravity see
\cite{Burgess:2003jk, Donoghue:2012zc, Goldberger:2007hy, Burgess:2019EFTB}).
The broad category of theories whose low-energy effects are described by EFTs
must only: ($a$) satisfy the assumption that it not introduce new degrees of
freedom at the length scales (call them $r$) relevant for the observations; and
($b$) have a significant hierarchy, $\ell \ll r$, between $r$ and the
new-physics scales (collectively represented by $\ell$). The main drawback of
EFTs relative to a model-by-model approach is a comparative lack of
predictability; an EFT's effective interactions are all {\it a-priori}
independent of one another, while the predictions of any particular model would
express them in terms of the smaller set of model parameters.

Because these two approaches have complementary strengths it is usually the
most informative when both are deployed. It is the comparison between specific
model predictions and generic EFT expectations that most quickly focuses
attention on what the most promising directions consistent with observations
might be.   

\subsection*{Black holes {\it v.s.} compact sources: a new type of EFT}

Although EFTs can play an important role in the post-Newtonian regime of
black-hole mergers \cite{Goldberger:2004jt}, comparatively little work as been
done computing the predictions of EFTs in the strong-gravity, near-horizon
regime for black holes. There are two main reasons for this. 

The first reason is a practical one: EFTs typically involve a local expansion
in powers of derivatives (both curvatures and derivatives of any other fields),
like
\be \label{Sbulk}
  S_{\rm eff} = - \int \exd^4x \, \sqrt{-g} \; \left[ - \frac{M_p^2}{2} \, \cR
  + (\partial \phi)^2 + \frac14 \, F_{\mu\nu} F^{\mu\nu} + c_1 \, \cR^2 + c_2
  \cR \, F_{\mu\nu} F^{\mu\nu} + c_3 (\partial \phi)^4 + \cdots \right] \,,
\ee
where $\cR$ is the Ricci scalar, $F_{\mu\nu}$ is a gauge field strength (such
as for electromagnetism) and $\phi$ is any scalar fields that might be
entertained at the energies of interest. The ellipses here involve all possible
powers of curvatures and fields and their derivatives. These actions inevitably
involve the presence of higher-derivative interactions. Techniques are only now
being developed \cite{Allwright:2018rut, Cayuso:2017iqc, Okounkova:2019dfo,
Okounkova:2019zjf} to handle efficiently such interactions in the
strong-gravity, near-horizon regime. The problem is that higher derivative
interactions introduce spurious solutions that have nothing to do with the full
theory's low-energy limit. These are known not to cause problems of principle
for EFTs (since the spurious solutions do not arise at fixed orders in the
expansion in powers of $\ell/r$ --- see {\it e.g.} the discussions in
\cite{Burgess:2014lwa, Solomon:2017nlh, Burgess:2019EFTB}) this is only cold
comfort for numerical calculations, where it is difficult in practice to
separate the spurious evolution from real predictions of the low-energy regime. 

The second obstruction to using EFT methods in the strong-gravity, near-horizon
regime arises because many ideas about small-distance extensions of GR are
quite speculative. Although often well-motivated, the resulting theoretical
frameworks are usually insufficiently developed to make predictions that are
precise enough for comparison with observations. For instance, in the
strong-gravity regime relevant to black holes, some such theories propose
qualitative changes like the appearance of an enormous number of new degrees of
freedom as one approaches the black hole horizon \cite{Almheiri:2012rt,
Susskind:2012uw, Lunin:2001jy, Mathur:2005zp}, but without a concrete framework
that allows phenomenological testing. It is usually assumed that these theories
only deviate from GR very close to the would-be event horizon, in order not to
alter the current understanding of the astrophysics of accretion disks.

As pointed out in \cite{Burgess:2018pmm}, theories that only modify GR in the
near-horizon regime can be systematically confronted with observations using a
slightly different EFT approach to modifications of GR.  This approach exploits
an expansion in powers of $\ell/r$ where $\ell$ characterizes the length scale
over which modifications are significant and $r$ is a measure of the distance
scale of interest for the observations. (For the applications of this paper the
distances of interest prove to be of order the horizon size, $r \sim r_\ssH$.)
It is a special case of a more general framework \cite{Burgess:2016lal,
Burgess:2016ddi, Burgess:2017ekj, Plestid:2018qbf} that captures how boundaries
or compact objects affect their larger surroundings. For such theories the
implications of modifications can be explored using an effective action that is
localized near the horizon, expanded in powers of ordinary fields and their
derivatives. Because EFTs capture the the low-energy implications of {\it any}
possible ultraviolet (UV) extension, confronting the EFT with observations
allows the implications of these theories to be explored even though a detailed
UV completion is not yet known.

In this paper, we extend the application of these techniques beyond the
applications to LIGO considered in \cite{Burgess:2018pmm} to include recent
results from the EHT~\cite{Akiyama:2019cqa}. EHT observations of the event
horizon are consistent with M87 being a supermassive black hole described by
the Kerr metric, {\it i.e.}~a black hole that is purely described by its mass
and spin as in GR. We argue that this agreement can be used to constrain
theories of modified gravity and in particular those whose effects are
localized near the event horizon. 

\subsection*{Reflection, transmission and experiments}

As shown in \cite{Burgess:2018pmm}, at lowest nontrivial order in $\ell/r$ the
observational implications of theories with exclusively near-horizon deviations
from GR can be described in terms of a near-horizon reflection coefficient,
$R$, where $|R|^2$ represents the reflection probability for an inward-directed
wave\footnote{More precisely, partially reflecting boundary conditions are
imposed, say, on a surface just outside the horizon, such as at radial position
$r = r_\ssH +\epsilon$.  The discussion below reviews the arguments of
\cite{Burgess:2018pmm} as to why nothing physical depend on the precise choice
of $\epsilon$.} `at' the horizon. GR emerges as the limit $R \to 0$, where
every wave is purely infalling sufficiently near the horizon. For more general
$R$ a nonzero fraction of an incoming wave is reflected at the would-be event
horizon, and it is the implications of these that provide potential new-physics
signatures.\footnote{The replacement of UV physics with boundary conditions
also applies elsewhere in physics, such as the influence of nuclear structure
on atomic energy levels (where it is the small ratio, $r_\ssN/a_\ssB$, of
nuclear to atomic size that controls the EFT). The observation that nuclear
properties affect atomic energy levels only through such boundary conditions
allows them to be computed fairly efficiently \cite{Burgess:2017mhz}.} The
relevance of near-horizon reflection to such theories is in any case intuitive,
and the hypothesis of nonzero $R$ was earlier used\footnote{Earlier work tends
not to distinguish the scales $\epsilon$ and $\ell$, which in general can be
very different. For applications to the effects of nuclear structure on atomic
energy levels the RG-invariant scale is $\ell \sim (Z\alpha)^2 r_\ssN$; much
smaller than nuclear size, $r_\ssN$, despite choosing boundary conditions
outside the nucleus: $\epsilon > r_\ssN$ \cite{Burgess:2017ekj,
Burgess:2017mhz}.} to identify interesting signatures such as for
gravitational-wave echoes~\cite{Cardoso:2016rao, Cardoso:2016oxy,
Abedi:2016hgu, Maselli:2017tfq, DAmico:2019dnn}.

In the context of the EHT, the deviations from GR that are captured by a
reflection coefficient can be constrained in the following way: the EHT
observes radio waves that are emitted as synchrotron radiation from the
relativistic electrons in the hot magnetized optically thin accretion disk of
M87. If there is a sufficient amount of reflection near the horizon the image
as seen by the EHT deviates from the predictions of GR: the radio waves emitted
by the accretion disk that fall into the black hole in vanilla GR become
instead partly reflected and escape. Some of these waves would then survive to
be seen by the EHT. In principle, sufficiently large deviations between the
observed EHT picture and what is consistent with a vanilla black hole can be
excluded. This type of logic has been used to constrain several theories of
modified gravity, such as naked singularities~\cite{Bambi:2019tjh,
Shaikh:2019hbm}, extra dimensions~\cite{Vagnozzi:2019apd, Banerjee:2019nnj},
scalar hair~\cite{Roy:2019esk, Cunha:2019ikd, Banerjee:2019xds}, and magnetically charged
black holes~\cite{Allahyari:2019jqz}. The focus of this work is to follow
\cite{Burgess:2018pmm} and show how to constrain \textit{all} UV modifications
of GR whose implications are localized near the horizon, and to use the EFT
framework to quantify the theoretical error that is involved in making
potentially-observable predictions.

As a proof of concept, we implement the program of constraining different
reflection coefficients in a particularly simple way. As is standard in these
calculations, we imagine a near-horizon reflection coefficient to have been
specified on a surface just outside the horizon.\footnote{In practical examples
we choose this to be a surface of fixed coordinate radius at $r = r_\ssH +
\epsilon$.} (Although much of the literature specializes to Schwarzschild
geometries, we here consider a reflection coefficient --- in both spherically
symmetric and direction-dependent versions --- specified on a surface just
outside a Kerr black-hole horizon.\footnote{ One might worry that physical
results depend on the precise position of this surface, but it turns out the
precise position of this surface is irrelevant because physical observables do
not depend on it. This does not however preclude them from depending on the
physical (but distinct) length-scale $\ell$~\cite{Burgess:2018pmm}. }) Although
some earlier workers perform their phenomenological analyses using relatively
extreme reflection probabilities, like $|R| = 1$ --- {\it i.e.}~perfect
reflection --- we instead keep this parameter general since this allows us to
quantify how big $R$ can be and remain consistent with observations. We show
below that (for electromagnetic waves, at least) extreme choices like $|R|=1$
are very likely already ruled out by the EHT.

\subsection*{Modelling the compact object's environment}

In principle, to constrain the existence of (or discover) new physics one must
simply compute the image expected with GR and the image expected in the
presence of near-horizon reflection, and ask whether these can be distinguished
by EHT observations. In practice, the great complication comes from modelling
the compact object's environment, since this is ultimately the source of the
light being imaged. For M87 this environment is believed to be an optically
thin accretion disk~\cite{Narayan:1994xi, Narayan:1994is, Reynolds:1996is},
whose detailed modelling is required when constructing and interpreting the
EHT images \cite{Akiyama:2019bqs, Akiyama:2019fyp, Akiyama:2019eap}.

Rather than providing a similarly detailed model of this environment, we here
take a simpler approach that builds on the fact that the observations seem in
first approximation to be well-described by GR. We obtain our constraint by
starting with an EHT image and tracing the observed light rays back to the
vicinity of M87 under the assumption that they pass through the vanilla Kerr
black-hole geometry predicted by GR. We ignore light scattering when doing so
because the black-hole environment is optically thin. This ray-tracing exercise
allows us to map the observed light intensity back to a light-intensity map on
any particular reference surface that surrounds the compact object. To see the
effects of reflection we then re-propagate these light rays back out to an
observer at infinity,\footnote{Strictly speaking, we actually re-propagate rays
back in to the compact object from infinity to find where on the reference
surface they arrive.} again using the Kerr geometry, but this time assuming a
nonzero near-horizon reflection coefficient, $R$, on a surface very near to the
position of the black-hole horizon. Sufficiently much reflection of this type
alters the intensity map seen at infinity (because, for example, light that
otherwise might have fallen into the black hole now gets reflected and escapes
to infinity), causing it to differ from the starting EHT image. The picture of
the reflected rays is then added to the original EHT picture and if the
combined picture is inconsistent with the diagnostics established by the EHT
collaboration~\cite{Akiyama:2019bqs, Akiyama:2019fyp, Akiyama:2019eap}, this
particular reflection coefficient can be excluded. 

There are three key assumptions in this reasoning. The first is that the medium
is optically thin, so that ray intensity does not vary appreciably while en
route to the observer. The second is that any particular element of the
reference surface radiates equal intensity into all directions, so that the
intensity of the reflected rays depends only on where on the reference surface
it starts (together with the precise value of the reflection coefficient). The
third assumption is that the effects of reflection are small; there is not
enough of it to feed back on the environment and alter the intensity
distribution inferred in the absence of reflection. Our explicit ray-tracing
code is available online at 
\url{https://github.com/mrummphys/EventHorizonTelescope}.

We find that we can typically constrain reflection coefficients to be less than
of order $1-10$\%, which corresponds to to constraining the EFT scale to
$\ell < (7-70) \mu m$. These constraints only weakly depend on the so-far poorly
constrained spin of the compact object and the angular dependence assumed for
$R$. The most constraining image statistics turn out to be the deviation from
circularity of the EFT image's ring-like structure and, to a lesser extent, its
fractional central brightness ({\it i.e.}~the ratio of intensity in the center
of the image to the intensity in the ring-like structure). 

We emphasize that our analysis presented here is based only on a digitized
image of the EHT observations. One should be able to improve these bounds
considerably using the original (and future) EFT dataset, and by incorporating
reflection directly into the relativistic magnetohydrodynamic (GRMHD)
simulations used by~\cite{Akiyama:2019fyp} to model the compact object's
environment. Although these extra steps can and should be done, our analysis
both provides a demonstration of concept and a relatively simple way to
estimate the size of EHT constraints on classes of deviations from GR. We hope
to address some of these issues in future work.

\subsection*{A road map}

This paper is structured as follows: in Section~\ref{sec:refl}, we introduce
reflection coefficients as described by the EFT we use for near-horizon
physics. In Section~\ref{sec:EHTbackground}, we introduce the EHT observations
and image analysis techniques used by the EHT collaboration followed by a
description of our ray tracing and image creation ethology in
Section~\ref{sec:methology}. The results are discussed in
Section~\ref{sec:results}, followed by our conclusions in
Section~\ref{sec:conclusions}. Appendix \ref{appendix:Rh} recaps the connection
between $R$ and the fundamental-physics scale $\ell$.

\section{The near-horizon EFT analysis}
\label{sec:refl}

An interesting class of approaches to black-hole information-loss problems
argue that deviations from GR arise only as an observer approaches the event
horizon~\cite{Almheiri:2012rt, Susskind:2012uw, Lunin:2001jy}. For example, one
proposal in this category argues black holes are not really black; instead they
just involve a very large number of new microscopic degrees of freedom whose
huge degeneracy is related to the enormous black hole entropy. Although it is
usually hard to make precise predictions in these proposals about the
properties of astrophysical black holes, these new degrees of freedom are
expected to become accessible only close to where the would-be black hole
horizon forms~\cite{Lunin:2002qf, Mathur:2005zp}. 

Having any hypothetical new physics be localized to lie within a distance
$\ell$ of the horizon, with $\ell$ much smaller than the horizon size itself,
$\ell \ll r_\ssH$, would be in any case also attractive since such a
construction would have been relatively easy to miss until the present. Both of
these reasons argue for quantifying the observational implications of such
theories for the new observational windows into the near-horizon regime. A
low-energy EFT approach would be particularly valuable since its validity can
be much broader than any particular (as yet, possibly, ill-defined)
UV completion. Because they involve systematic low-energy expansions, EFT
methods also lend themselves to explicitly quantifying any theoretical
uncertainties intrinsic to the predictions. We now sketch the construction of
the EFT for exotic near-horizon physics, and reproduce the argument of
ref.~\cite{Burgess:2018pmm} that its leading effect is to give a reflection
coefficient (and, sometimes, a damping time) for any particular mode. 

\subsection{EFTs and near-horizon boundary conditions}

The starting point is the effective action. When $\ell$ is much smaller than
the scales of practical interest (such as the horizon size $r_\ssH$) then it is
useful to integrate out all of the new physics associated with any new degrees
of freedom localized near $\Sigma$. Once this is done the influence of this new
physics is captured by a component of the effective action that is localized on
a surface very near the horizon. For example, for a single complex Klein-Gordon
scalar field in a black-hole background this leads to an action of the form $S
= S_\ssB + S_{\rm hor}$ where
\be
  S_\ssB = -  \int \exd^4x \, \sqrt{-g} \Bigl[ g^{\mu\nu} \partial_\mu \phi^*
  \, \partial_\nu \phi + m^2 \phi^* \phi + \cdots \Bigr]
\ee
and
\be \label{horaction}
   S_{\rm hor} = - \int_\Sigma \exd^3 x \; \sqrt{- \gamma}\;  \Bigl[h_0 + h_1
   \, \phi^* \phi + \cdots \Bigr] \,,
\ee
where the integration in \pref{horaction} is over a time-like surface,
$\Sigma$, defined by $x^\mu = y^\mu(\sigma^a)$ just outside the horizon ({\it
e.g.}~perhaps $r = r_\ssH + \epsilon$ for a Schwarzschild black hole), and
$\gamma_{ab} = g_{\mu\nu} \partial_a y^\mu \, \partial_b y^\nu$ is the induced
metric on $\Sigma$. 

For both $S_\ssB$ and $S_{\rm hor}$ the ellipses denote all possible local
combinations involving more powers of both the field and its derivatives. The
more fields or derivatives appearing in such terms, the more suppressed by
powers of the microscopic length scale $\ell$ their effective couplings must
be. For instance, for a canonically normalized scalar like $\phi$ the coupling
$h_1$ in \pref{horaction} has dimensions (length)${}^{-1}$. An effective
interaction in $S_b$ proportional to $h_n (\phi^* \phi)^n$, on the other hand,
would have a coupling $h_n$ with dimension (length)${}^{2n-3}$.  The reality of
the effective coupling $h_n$ is related to probability conservation. If overall
probability is conserved at the horizon, $h_n$ is real but if not -- as in the
black hole case -- $h_n$ is complex with the imaginary part of $h_n$
quantifying probability loss/absorption of the system~\cite{Burgess:2018pmm}.

Effective couplings (like $h_1$) appearing in $S_b$ affect the dynamics of
$\phi$ away from $\Sigma$ only through the boundary condition that they
contribute near $\Sigma$. This can be seen most easily in the semiclassical
limit when examining the saddle point for functional integrations over $\phi$
(both on and off the surface $\Sigma$). Variations off the surface (in the
`bulk') reproduce the classical field equation, $(-\Box + m^2) \phi = 0$, while
demanding a saddle point for $S_\ssB + S_{\rm hor}$ right at $\Sigma$ give the
boundary condition
\be \label{BC1}
  \Bigl( n_\mu g^{\mu\nu} \partial_\nu\phi  + h_1 \phi +\cdots \Bigr)_\Sigma = 0 \,,
\ee
where $n_\mu$ is the unit outward-pointing normal to $\Sigma$ and ellipses
denote the influence of any terms hidden in the ellipses of
eq.~\pref{horaction}. 

This boundary condition \pref{BC1} can be interpreted in two complementary
ways. First, the value of $h_1$ dictates the normal derivative of $\ln \phi$ on
$\Sigma$, which in turn dictates one of the integration constants found when
integrating the Klein Gordon equation in the bulk. Since these integration
constants in turn control the relative size of ingoing and outgoing modes, a
straightforward argument shows that knowledge of the pair $(h_1, \epsilon)$ is
equivalent to knowledge of the reflection coefficient $R[h_1,\epsilon]$.
Although this makes it seem as if $R$ depends on $\epsilon$, this is not really
true because the value of $h_1$ found by integrating out UV physics is itself
$\epsilon$-dependent, in just such a way that $(\exd/\exd
\epsilon)R[h_1(\epsilon), \epsilon] = 0$. That is, the functional form of
$h_1(\epsilon)$ is precisely what is required to ensure that $R$ does not
depend on $\epsilon$. 

This is a special case of a general EFT argument: the regularization scale
$\epsilon$ drops out of observables because it is absorbed into a
renormalization of $h_1(\epsilon)$, defining an RG flow $h_1(\epsilon/\ell)$,
for some RG-invariant scale $\ell$. This provides the second, complementary,
way to read eq.~\pref{BC1}: because it holds for {\it any} value of $\epsilon$
it remains true even after differentiation with respect to $\epsilon$. If this
is done with the reflection coefficient held fixed it provides the precise form
for $h_1(\epsilon)$ required to ensure that $R$ is $\epsilon$-independent.
This allows $R[h_1(\epsilon/\ell),\epsilon]$ to be traded for a more
informative relationship $R = R(\ell/r_\ssH)$ between $R$ and the physical
RG-invariant length-scale $\ell$ (given in more detail in
Appendix~\ref{appendix:Rh}.) Upper limits to $|R|$ in this way turn into upper
limits on $\ell/r_\ssH$. 

For black holes the infrared fixed point of the RG flow one finds in this way
for $h_1$ is complex and corresponds to the choice of purely infalling boundary
conditions at the horizon, {\it i.e.}~to vanilla GR~\cite{Burgess:2018pmm}.
Details of this construction are given in~\cite{Burgess:2016lal,
Burgess:2016ddi, Burgess:2017ekj, Burgess:2018pmm} and so are not repeated
here. Instead, for the purposes of EHT phenomenology we simply imagine the
reflection coefficient $R$ to have been specified.

\subsection{Reflection Coefficients}

Given that a class of modifications of GR can be described by replacing the
perfect-infall boundary condition of GR with a mixture of incoming and outgoing
wave near the horizon, we next turn to what potentially observable consequences
this might have. As has been pointed out, partial reflection of incoming
gravitational waves near the horizon can lead to a series of echoes that might
be observable at LIGO~\cite{Cardoso:2016rao, Cardoso:2016oxy, Cardoso:2019rvt,
Abedi:2016hgu, Maselli:2017tfq}. In this paper, we instead explore the
implications of this kind of boundary condition for electromagnetic waves as
probed by the EHT.

To study reflection we need to specify a reflection coefficient $R$ at a
surface $\Sigma$, which for simplicity we choose to be a surface of fixed
coordinate radius, $r = r_\ssR = r_\ssH + \epsilon$ with $\epsilon \ll
r_\ssH$. As discussed above, the value of $r_\ssR$ is not physically relevant
as $R$ is RG invariant. In specifying $R$ we select a particular representation
of an RG flow representing a particular UV completion. We will look into three
different kind of possibly angular dependent reflection coefficients inspired
by a multipole expansion:
\begin{align}
  \begin{aligned}
    \label{eq:Rl_def}
    &l=0:\quad R(\theta, \phi) = R_0\,,\\
    &l=1:\quad R(\theta, \phi) = R_0\, |\cos(\theta) |\,,\\
    &l=2:\quad R(\theta, \phi) = R_0\, |\sin(\theta) \cos(\theta) |\,,\\
  \end{aligned}
\end{align}
where the higher multipoles $l=1,2$ introduce a dependence on the polar angle
$\theta$. One might expect more complicated angular dependence in a particular
UV modification but~\eqref{eq:Rl_def} is a starting point for the simplest
reflection coefficients one might expect. For a discussion how the angular
dependence of $R$ is related to the angular dependence of the
point-particle-effective-field-theory (PPEFT) coupling describing the black
hole, see~\cite{Burgess:2018pmm} and Appendix~\ref{appendix:Rh}. Finally, $R$
might also be frequency dependent in a generic UV modification, i.e.~a waves'
incoming frequency might be different from the reflected outgoing waves'
frequency due to energy absorption/emission by the black hole object. A
backreaction on the background geometry may occur close to the horizon, see
e.g.~\cite{Carballo-Rubio:2018vin,Chen:2019hfg,Cardoso:2019rvt}. This would
require modelling additional effects which we do not take into account in this
work and instead focus on the simpler scenario described by~\eqref{eq:Rl_def}.

\section{EHT observations}
\label{sec:EHTbackground}

The EHT~\cite{Akiyama:2019cqa} is a very long baseline interferometry
experiment that measures radio brightness distributions at a wavelength of 1.3
mm on the sky. Its unprecedented angular resolution allows to resolve
structures of angular scales of $\mathcal{O}(\mu{\rm as})$. This makes it
possible for the first time to resolve event-horizon-scale physics of
supermassive black holes that are relatively nearby and/or active, i.e.~bright.
In particular, the EHT collaboration was able to reveal the shadow caused by
gravitational light bending and photon capture at the event horizon of the
supermassive black hole at the center of the giant elliptical galaxy M87 at a
distance of 16.8 Mpc~\cite{2018ApJ...856..126C}. The images of M87 show a
ring-like structure with a diameter of about 40 $\mu$as with a central
brightness depression due to the event horizon of the black hole. The ring
brightness is asymmetric which can be explained by relativistic beaming of the
photons emitted as synchrotron radiation from the plasma rotating in the
accretion disk at close to the speed of light. The angular size of the ring is
directly related to the mass (and to a lesser extend spin) of the supermassive
black hole and is estimated by the EHT collaboration as $M = (6.5\pm0.7)\cdot
10^9 M_\odot$~\cite{Akiyama:2019cqa}. There is a slight tension with estimates
of the mass via stellar dynamics at $M = 3.5^{+0.9}_{-0.3}\cdot 10^9
M_\odot$~\cite{Walsh:2013uua}.

The images of the EHT collaboration are consistent with a shadow of
a rotating Kerr black hole in general relativity. The EHT observations offer a
new unique opportunity to test near horizon gravitational physics, such as
extensions to general relativity described in Section~\ref{sec:refl}.

In order to constrain modified theories of gravity we first need an image of
the EHT observations that we can compare the modified gravity images to. Since
the EHT observations are consistent with general relativity, i.e.~a Kerr black
hole, a modified theory of gravity can be excluded if its image is too
dissimilar from the EHT/Kerr image. To determine if two images are dissimilar
we invoke the image analysis variables that we will introduce in
Section~\ref{sec:ImageAnalysis}.

\begin{figure}[!htb]
\centering
\includegraphics[width=.7\textwidth]{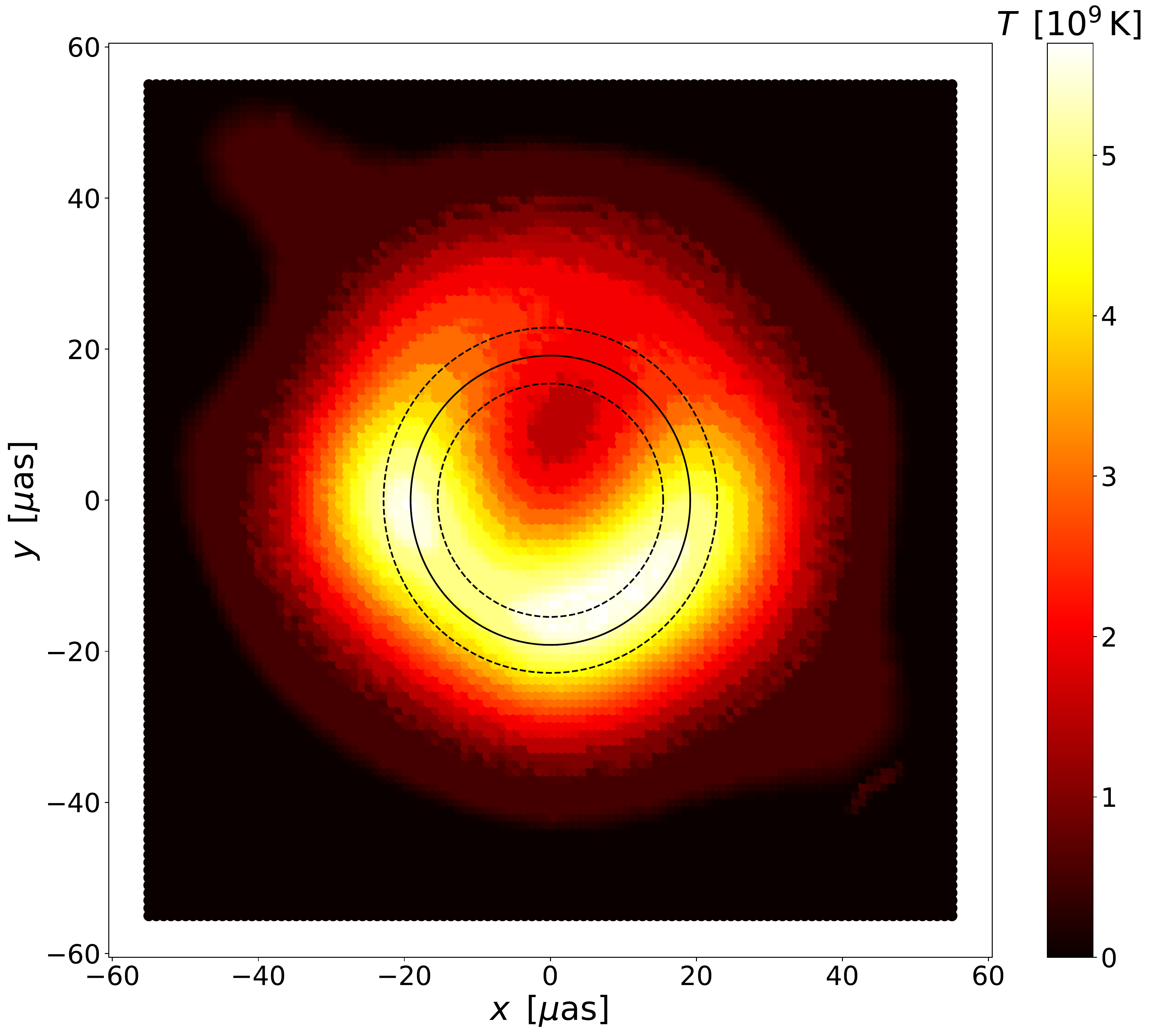}
\caption{\label{fig:EHT_digitized} The digitized EHT image. The solid black
  line is half a ring diameter $d/2$ from the image center where $d$ is defined
  in eq.~\eqref{eq:d}. The dashed black lines are at radii $(d-\sigma_d)/2$ and
  $(d+\sigma_d)/2$, respectively where $\sigma_d$ is defined
  in~\eqref{eq:sigmad}.}
\end{figure}

As we do not have access to the EHT data, we digitize one of the EHT images.
Since this is a proof-of-concept study a digitized version will be good enough
for our purposes however we generally expect our bounds derived in
Section~\ref{sec:results} to be more sensitive by a factor of a few with the
original data. A generic representative of the different imaging methods
\texttt{DIFMAP}, \texttt{eht-imaging} and \texttt{SMILI}
(see~\cite{Akiyama:2019fyp}) of the EHT is Figure~15 in \cite{Akiyama:2019fyp}
which is an average of these three methods observed on April 11 2017. To
digitize the image, we use the colour code provided in the figure to find all
pixels with brightness temperature $T$ in a linear sample between 0 and $6\cdot
10^9$ K with $5\cdot 10^8$ K step size. Next, we create a pixel grid of size
-55 to 55 $\mu$as with pixel size of 1 $\mu$as. Each pixel is assigned an
intensity via a 2D interpolation function from the above points of the original
image. The obtained image is shown in Figure~\ref{fig:EHT_digitized}.

\subsection{Image analysis}
\label{sec:ImageAnalysis}

Here, we give a brief review of the image diagnostics the EHT collaboration
uses to characterize the ring like structure in an image. By comparing these
diagnostics of an image from modified gravity to the observed EHT image we can
determine if a theory is either excluded or consistent with the experiment.

Following Section~9 in~\cite{Akiyama:2019fyp}, we first have to find the center
of the ring like structure which is the position $(x_0, y_0)$ in the image that
minimizes the normalized peak dispersion, i.e.
\begin{equation}
  (x_0, y_0) = {\rm argmin} \left[ \frac{\sigma_{\bar r}(x, y)}{\bar r_{\rm
  pk}(x,y)}\right]_{(x, y)}\,,
\end{equation}
where $\bar r_{\rm pk}(x,y)$ and $\sigma_{\bar r}(x, y)$ are the mean and
standard deviation of the peak brightness
\begin{equation}
  r_{\rm pk}(\theta; x, y) = {\rm argmax}_r \left[I(r,\theta;x,y) \right]\,,
\end{equation}
with respect to $\theta$. $I(r,\theta;x,y)$ is the intensity of the image at
pixel location defined by $(r, \theta)$ coordinates with respect to an origin
at $(x, y)$. In practice, $\theta$ and $r$ are linearly sampled over a
discrete set of values between 0 and $50\,\mu$as and 0$^\circ$ and 360$^\circ$
respectively while we select $x$ and $y$ values each from 20 equally spaced
values between $-8$ and $8\,\mu$as. In order to avoid spurious detection, \
\cite{Akiyama:2019fyp} limits the peak finding algorithm to pixels that have
at least $95\%$ of the image's peak brightness. Since our digitized version of
the EHT observation has about three times worse brightness resolution we loosen
this constraint to at least $60\%$ of peak brightness instead.

Once the image center is found the measured diameter and its uncertainty are
defined as the mean
\begin{equation}
  \label{eq:d}
  d = 2\bar r_{\rm pk}(x_0,y_0)\,,
\end{equation}
and standard deviation
\begin{equation}
  \label{eq:sigmad}
  \sigma_d = 2\sigma_{\bar r}(x_0,y_0)\,.
\end{equation}
We will suppress the $(x_0, y_0)$ argument in the following for all quantities.
The deviation from circularity is defined as
\begin{equation}
  \Delta_c = \frac{\sigma_d}{d}\,,
\end{equation}
which will be a key quantity to constrain modified theories of gravity
effectively described by a reflection coefficient in Section~\ref{sec:results}.

The ring width is defined as
\begin{equation}
  \label{eq:width}
  w = \langle {\rm FWHM \left[I(r, \theta) - I_{\rm floor}
  \right]}\rangle_\theta\,,
\end{equation}
where FWHM is full width half maximum over a radial profile at given angle
$\theta$. $I_{\rm floor} = \langle I(r_{\rm max} = 50\mu{\rm as},
\theta)\rangle_\theta$ is subtracted from the intensity profile to avoid
introducing bias between different images. While~\eqref{eq:width} defines the
mean we can also calculate $\sigma_w$ as the standard deviation of the set of
FWHMs in~\eqref{eq:width}.

The ring orientation angle is defined as
\begin{equation}
  \label{eq:eta}
  \eta = \left\langle {\rm arg} \left[ \int_0^{2\pi} I(\theta) e^{i\theta}
  \exd\theta\right] \right\rangle_{r \in [r_{\rm in}, r_{\rm out}]}\,,
\end{equation}
i.e.~the mean over radii between $r_{\rm in} = (d-w)/2$ and $r_{\rm out} =
(d+w)/2$ of the first angular mode of the angular profile $I(\theta)$.
$\sigma_\eta$ is defined as the standard deviation over the set of radii
in~\eqref{eq:eta}.

The degree of azimuthal asymmetry is defined as the mean
\begin{equation}
  A = \left\langle \frac{\int_0^{2\pi} I(\theta) e^{i\theta}
  \exd\theta}{\int_0^{2\pi} I(\theta)
  \exd\theta} \right\rangle_{r \in [r_{\rm in}, r_{\rm out}]}\,,
\end{equation}
with $\sigma_A$ defined as the corresponding standard deviation. $A$ takes
values between 0 and 1 and can be interpreted as how evenly the brightness in a
ring is distributed over the azimuth angle. Perfect azimuthal symmetry
corresponds to $A=0$ while a delta function concentrating all brightness at one
particular angle corresponds to $A=1$.

The fractional central brightness is defined as
\begin{equation}
  f_c = \frac{\langle I(r,\theta)\rangle_{\theta, r\in[0, 5\mu{\rm
  as}]}}{\langle I(d/2,\theta)\rangle_{\theta \in[0, 2\pi]}}\,,
\end{equation}
i.e.~the ratio of the mean of the intensity within a disk of radius 5 $\mu$as
and the mean brightness around the ring.

For our digitized EHT observation on April 11 we find the image characteristics
listed in Table~\ref{tab:image_diagnostics} compared to those listed in
\cite{Akiyama:2019fyp}.
\begin{table}[!htb]
 \centering
 \begin{tabular}{||l c c c c c c||}
 \hline
    & $d$ ($\mu$as) & $w$ ($\mu$as) & $\eta$ ($^\circ$) & $A$ &
   $\Delta_c$ & $f_c$ \\
 \hline\hline
   Digitized & $38.3 \pm 7.4$ & $28.4 \pm 7.4$ & $205.1 \pm
   83.7$ & $0.15 \pm 0.08$ & $0.19$ & $0.63$ \\
 \hline\hline
   \texttt{DIFMAP} & $40.7 \pm 2.6$ & $29.0 \pm 3.0$ & $173.3 \pm
   4.8$ & $0.23 \pm 0.04$ & $0.06$ & $0.5$ \\
 \hline
   \texttt{eht-imaging} & $41.0 \pm 1.4$ & $15.5 \pm 1.8$ & $168.0
   \pm 6.9$ & $0.20 \pm 0.02$ & $0.03$ & $0.04$ \\
 \hline
   \texttt{SMILI} & $42.3 \pm 1.6$ & $15.6 \pm 2.2$ & $167.6 \pm
   2.8$ & $0.22 \pm 0.03$ & $0.04$  & $6\cdot10^{-6}$ \\
 \hline
 \end{tabular}
 \caption{Image characteristics of our digitized image compared to the EHT
   image pipelines for the April 11 observation.}
 \label{tab:image_diagnostics}
\end{table}
Note that we use an overlayed/averaged version of \texttt{DIFMAP},
\texttt{eht-imaging} and \texttt{SMILI} and \texttt{DIFMAP} has the biggest
blur.\footnote{As referred to in \cite{Akiyama:2019bqs} the blur of an image
depends on the beam size it has been restored with. \texttt{DIFMAP} uses the
largest beam size of the three image pipelines at 20$\mu$as while
\texttt{eht-imaging} and \texttt{SMILI} use 17.1$\mu$as and 18.6$\mu$as,
respectively.} Hence, we mainly have to compare to the \texttt{DIFMAP} row in
Table~\ref{tab:image_diagnostics} as the averaged image is dominated by its
blur.

Our image diagnostics in Table~\ref{tab:image_diagnostics} are consistent with
those of~\cite{Akiyama:2019fyp} but our errors are larger as is to be expected
from a digitized image. For diameter $d$, width $w$ and azimuthal asymmetry
$A$ the uncertainties are a factor 2 - 3 larger than in \texttt{DIFMAP} while
for the orientation angle $\eta$ the error is much larger. We conclude that
our digitized image has about three times worse brightness resolution than the
EHT image.

For the deviation from circularity and the fractional central brightness the
EHT collaboration reports the upper bounds~\cite{Akiyama:2019fyp}
\begin{equation}
  \label{eq:EHT_upper_bounds}
  \Delta_c \lesssim 0.1 \qquad {\rm and} \qquad f_c \lesssim 0.5\qquad ({\rm
  EHT})\,,
\end{equation}
as a summarizing result from their different images and pipelines. From our
digitized image we can only derive slightly weaker upper bounds on these
quantities from Table~\ref{tab:image_diagnostics}:
\begin{equation}
  \label{eq:digitized_upper_bounds}
  \Delta_c \lesssim 0.2 \qquad {\rm and} \qquad f_c \lesssim 0.7\qquad ({\rm
  digitized})\,,
\end{equation}
where we have rounded up to the next full digit from the results in
Table~\ref{tab:image_diagnostics} to be conservative. Digitizing the
overlayed/averaged image gives us a conservative scenario in constraining
$\Delta_c$ and $f_c$ as these quantities are the least constrained in
\texttt{DIFMAP} which dominates the errors in this overlayed image.

We can now formulate precisely what it means for an image generated from a
modified gravity theory to be consistent with EHT observations: An image is
consistent if the image characteristics $d$, $w$, $\eta$ and $A$ are in
agreement with the values in the first row of Table~\ref{tab:image_diagnostics}
and the upper bounds in~\eqref{eq:digitized_upper_bounds} are not violated.
Otherwise it is inconsistent and the theory is excluded by the EHT observations.

\section{Methodology}
\label{sec:methology}

In order to constrain theories of modified gravity described by a reflection
coefficient we create images that are the sum of the original EHT image and
an image of rays reflected close to the horizon at $r_\ssR$. Therefore, we first
describe our ray tracing algorithm in Section~\ref{sec:raytracing} and then
explain how the images from modified gravity theories are created in
Section~\ref{sec:imagecreation}.

\subsection{Ray tracing}
\label{sec:raytracing}

We need capture the light rays that approach the black hole, get reflected at
$r_\ssR$ and finally arrive at the camera. In principal, one possibility would
be to consider all initial conditions (position and direction) in the vicinity
of the black hole and trace their evolution according to the Kerr geodesic
equation to see which light rays arrive at the camera and from which direction.
As this is practically impossible a much more efficient computational way to
tackle this problem is to trace the light ray backwards in time,
i.e.~\textit{ray tracing}~\cite{Kidder:2000yq, Vincent:2011wz, Muller2014,
Bohn:2014xxa, Riazuelo:2015shp, Cunha:2018acu}. In this approach, the initial
condition for the geodesic equation is the position and angle at which the
light ray arrives at the camera, in this case the EHT. Then, the light ray is
evolved backwards in time according to the Kerr geodesic equation. At some
point, it might hit the surface of reflection at $r_\ssR$ where it is reflected
radially, while the intensity of the ray is modified according
to~\eqref{eq:Rl_def}. From there, it evolves away from the black hole to a
place infinitely far away from the black hole. In the following, we describe
our procedure of solving the Kerr geodesic equation and its initial conditions.

The geodesic equation for the coordinates $x^\mu(\lambda)$ for an affine
parameter $\lambda$ parametrizing the geodesic can be obtained by varying the
line element in GR as
\begin{equation}
	\label{eq:geodesic}
	\frac{\exd}{\exd\lambda}\,\frac{\partial e}{\partial \dot x^\mu} =
	\frac{\partial e}{\partial x^\mu}\,,
\end{equation}
where $\dot x^\mu \equiv \exd x^\mu / \exd\lambda$ and
\begin{equation}
	e = - g_{\mu\nu}\,\frac{\exd x^\mu}{\exd\lambda} \,
	\frac{\exd x^\nu}{\exd\lambda} = -g_{\mu\nu}\, \dot x^\mu\, \dot x^\nu\,.
\end{equation}
The Kerr metric in Boyer-Lindquist coordinates is
\begin{align}
\begin{aligned} \label{BoyerLind}
	\exd s^2 = &-g_{\mu\nu} \,\exd x^\mu \exd x^\nu\,,\\
	= &- \left(1 - \frac{r_s r}{\rho^2} \right)\exd t^2 - 
	\frac{2a r_s r \sin^2 \theta}{\rho^2} \exd t \exd\phi + 
	\frac{\rho^2}{\Delta} \exd r^2 +\rho^2 \exd\theta^2\\
	&+ \sin^2\theta \left[ \left(r^2 + a^2 \right) + 
	\frac{r_s r a^2 \sin^2\theta}{\rho^2}\right]\exd\phi^2\,,
\end{aligned}
\end{align}
where $M$ is the mass of the black hole, $a$ is the spin in units of $M$.
Furthermore (setting Newton's constant to unity)
\begin{align}
\begin{aligned}
 r_s &= 2M\,,\\
 \Delta &= r^2 - 2Mr+a^2\,,\\
 \rho^2 &= r^2 + a^2 \cos^2 \theta\,.
\end{aligned}
\end{align}
The horizon is located at
\begin{equation}
  \label{eq:r_horizon}
  r_\ssH = M + \sqrt{M^2 - a^2}\,.
\end{equation}

Eq.~\eqref{eq:geodesic} are four second order differential equations for four
variables $(t, r, \theta, \phi)$. By introducing the momentum variables
\begin{equation}
	p^\mu \equiv \dot x^\mu\,,
\end{equation}
these can be transformed into 8 first order differential equations. Using
energy conservation, the time variables $t$ and $p^t$ can be eliminated from
the equations, see e.g.~\cite{Chandrasekhar:1985kt}, and one is left with 6
first order differential equations for the variables $(r, \theta, \phi, p^r,
p^\theta, p^\phi)$
\begin{equation}
	\label{eq:geodesicwesolve}
	\dot x^i = p^i \qquad {\rm and} \qquad \dot p^i = F^i(x^j, p^j)\,,
\end{equation}
where the lengthy expressions $F^i(x^j, p^j)$ are listed in
Appendix~\ref{appendix:Kerrgeodesic}. We numerically
solve~\eqref{eq:geodesicwesolve} for the initial conditions discussed below
using the \texttt{scipy} package \texttt{odeint}.  To check our code,
we reproduce the results of~\cite{Bohn:2014xxa} for Kerr black holes.

The initial conditions for the backwards ray tracing are set at the camera. We
choose a Cartesian coordinate with the $x$-direction parametrizing the distance
to M87:
\begin{equation}
  \label{eq:cameraframe_initialconditions}
  \boldsymbol{x}_0 = \left( \begin{matrix}
           r_{\rm cam} \\
           (2h-1)\delta \\
           (2v-1)\delta \\
  \end{matrix} \right) \qquad {\rm and} \qquad
  \boldsymbol{p}_0 = \left( \begin{matrix}
           -E \\
           0 \\
           0 \\
  \end{matrix} \right)\,,
\end{equation}
where $r_{\rm cam}$ is the distance to M87, $\delta$ is half the angular size
of the image, $h,v \in [0,1]$ are $n_{\rm pixel} = 2\delta/\delta_{\rm pixel}$
equispaced variables each, where $\delta_{\rm pixel}$ is the pixel size,
indexing the pixel the light ray originates from. $E$ is the energy of the
ray/photon. Note that the rays are chosen to be parallel to the $x$-axis as the
distance to the black hole is much larger than its angular size, in fact about
$10^{10}$ larger as $\delta$ is of the order of 10th of $\mu$as. For the
numerics, this allows us to choose $r_{\rm cam}$ as a radius that is far enough
from the black hole for the metric to be sufficiently flat as opposed to the
huge physical distance to M87. In other words, the ray simply follows a straight
line trajectory parallel to the $x$-axis from this radius to the camera that we
do not have to include in the numerical calculations.

For our ray tracing algorithm we choose the following numerical values for the
above parameters:
\begin{equation}
    r_{\rm cam} = 100\mu{\rm as}\,,\qquad E = 1\,,\qquad \delta = 30\mu{\rm
    as}\,,\qquad \delta_{\rm pixel} = 1\mu{\rm as}\,.
\end{equation}
Note that the choice for $E$ here is somewhat arbitrary as the shape of the
trajectory of a ray is independent of its energy. Finally, we have to choose a
value for the black hole mass $M$ which we fix by the angular size of the ring
in the EHT image of M87. In~\cite{Akiyama:2019eap}, the EHT collaboration notes
that the ring diameter as defined in~\eqref{eq:d} typically arises at an
angular size that is 10\% larger than the photon ring, i.e.~at a radius of 21.1
$\mu$as for our digitized image, using the central value for $d$ from
Table~\ref{tab:image_diagnostics}. For a given black hole spin $a$ we can
calculate the size of the photon ring and choose $M$ such that it matches this
radius. This gives us values between $M = 5.7 \cdot 10^9 M_\odot$ $(a=0)$ and
$M = 6.1 \cdot 10^9 M_\odot$ $(|a|=0.94)$, using a distance of 16.8 Mpc to M87.

In order to numerically solve the differential
equations~\eqref{eq:geodesicwesolve} we need to transfer the initial position
and momentum in the camera frame to Boyer-Lindquist coordinates.  This is a two
step procedure: first, we need to
rotate~\eqref{eq:cameraframe_initialconditions} by the orientation angle of the
black hole which we set to $\theta_{\rm obs} = 17^\circ$~\cite{Walker:2018muw}.
The rotation matrix is:
\begin{equation}
  \boldsymbol{R}(\theta_{\rm obs}) =    \left( \begin{matrix}
    \sin \theta_{\rm obs} & 0 & - \cos \theta_{\rm obs}\\
    0 & 1 & 0 \\
    \cos \theta_{\rm obs} & 0 & \sin \theta_{\rm obs}\\
  \end{matrix} \right)\,.
\end{equation}
Second, we find the values for position and momentum in Boyer-Lindquist
coordinates by inverting the relations
\begin{equation}
  \label{eq:BL_initialconditions_x}
  \boldsymbol{R} \cdot \boldsymbol{x}_0 = 
   \left( \begin{matrix}
     r_{\rm cam} \sin \theta_{\rm obs} \\
     0 \\
     r_{\rm cam} \cos \theta_{\rm obs} \\
  \end{matrix} \right) = \left( \begin{matrix}
           \sqrt{r_0^2+a^2} \sin\theta_0 \cos\phi_0 \\
           \sqrt{r_0^2+a^2} \sin\theta_0 \sin\phi_0 \\
           r_0 \cos\theta_0\\
  \end{matrix} \right)\,,
\end{equation}
and
\begin{equation}
  \label{eq:BL_initialconditions_p}
   \boldsymbol{R} \cdot \boldsymbol{p}_0 = \left( \begin{matrix}
    \frac{r_0}{\sqrt{r_0^2+a^2}} p^r_0 \sin\theta_0 \cos\phi_0 
     + \sqrt{r_0^2+a^2} p^\theta_0 \cos\theta_0\cos\phi_0
     - \sqrt{r_0^2+a^2} p^\phi_0 \sin\theta_0\sin\phi_0 \\
    \frac{r_0}{\sqrt{r_0^2+a^2}} p^r_0 \sin\theta_0 \sin\phi_0 
     + \sqrt{r_0^2+a^2} p^\theta_0 \cos\theta_0\sin\phi_0
     + \sqrt{r_0^2+a^2} p^\phi_0 \sin\theta_0\cos\phi_0 \\
     p^r_0\cos\theta_0 - r_0 p^\theta_0 \sin\theta_0\\
  \end{matrix} \right)\,,
\end{equation}
for $(r_0, \theta_0,\phi_0)$ and $(p^r_0, p^\theta_0, p^\phi_0)$.

We implement ray reflection at
\begin{equation}
  r_\ssR = 1.2\, r_\ssH\,.
\end{equation}
As discussed in Section~\ref{sec:refl}, the choice for $r_\ssR$ is not physical
as only $R$ is physical and RG invariant. We have tested this explicityly
with various radii and shapes for the reflection surface and the image is
indeed independent of these features within the 1 $\mu$as resolution as long as
the hierarchy $\epsilon \ll r_\ssH$ is respected. The reflection condition is
imposed as a new initial condition for the differential equation solver if the
ray crosses the reflection surface $r_\ssR$:
\begin{equation}
  {\left(r, \theta, \phi, p^r, p^\theta, p^\phi \right)}_{r = r_\ssR} \rightarrow 
  {\left(r, \theta, \phi, - p^r, p^\theta, p^\phi\right)}_0
\end{equation}
i.e.~radial reflection.

\subsection{Image creation}
\label{sec:imagecreation}

\begin{figure}[!htb]
\centering
\includegraphics[width=.49\textwidth]{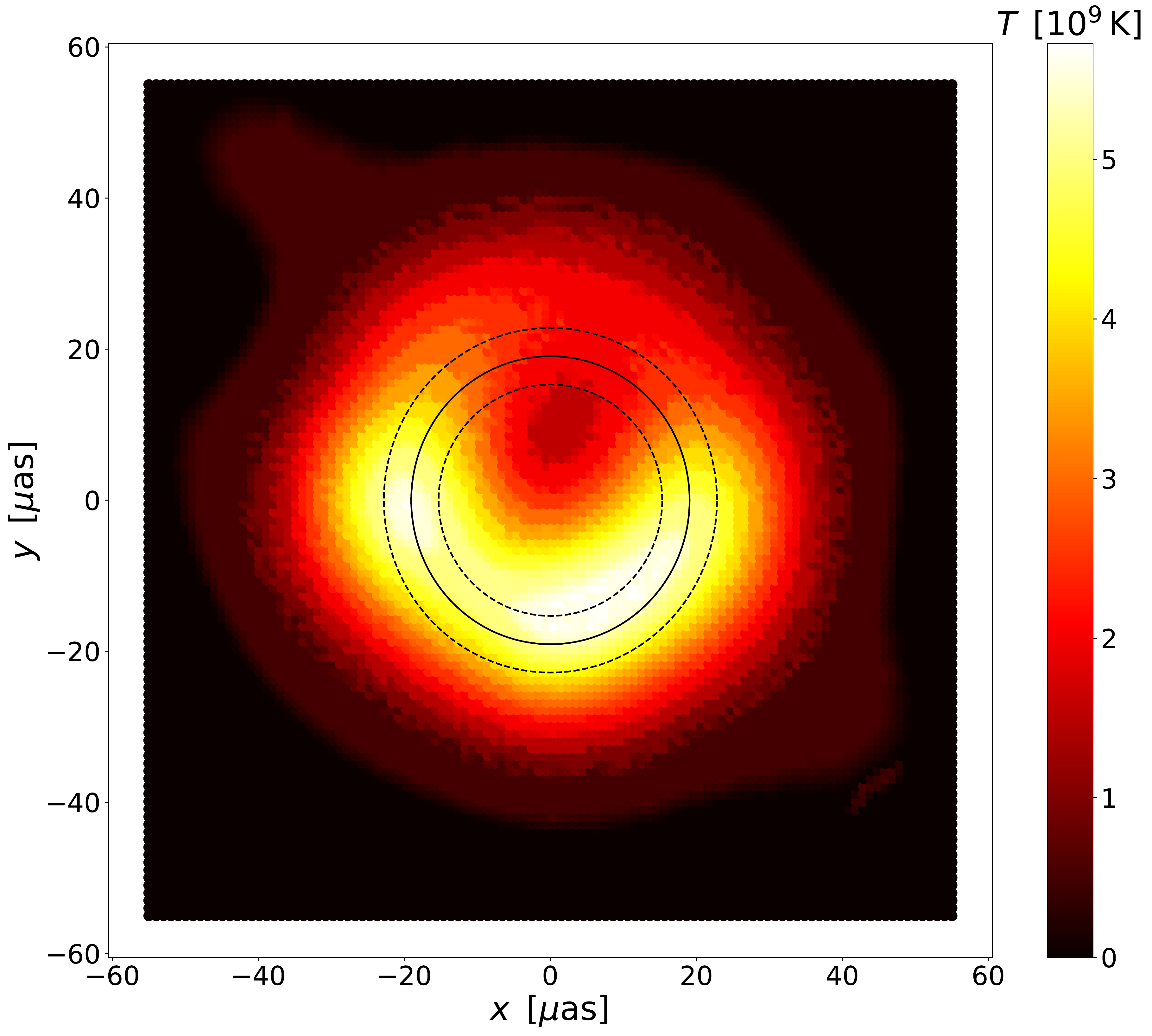}
\includegraphics[width=.49\textwidth]{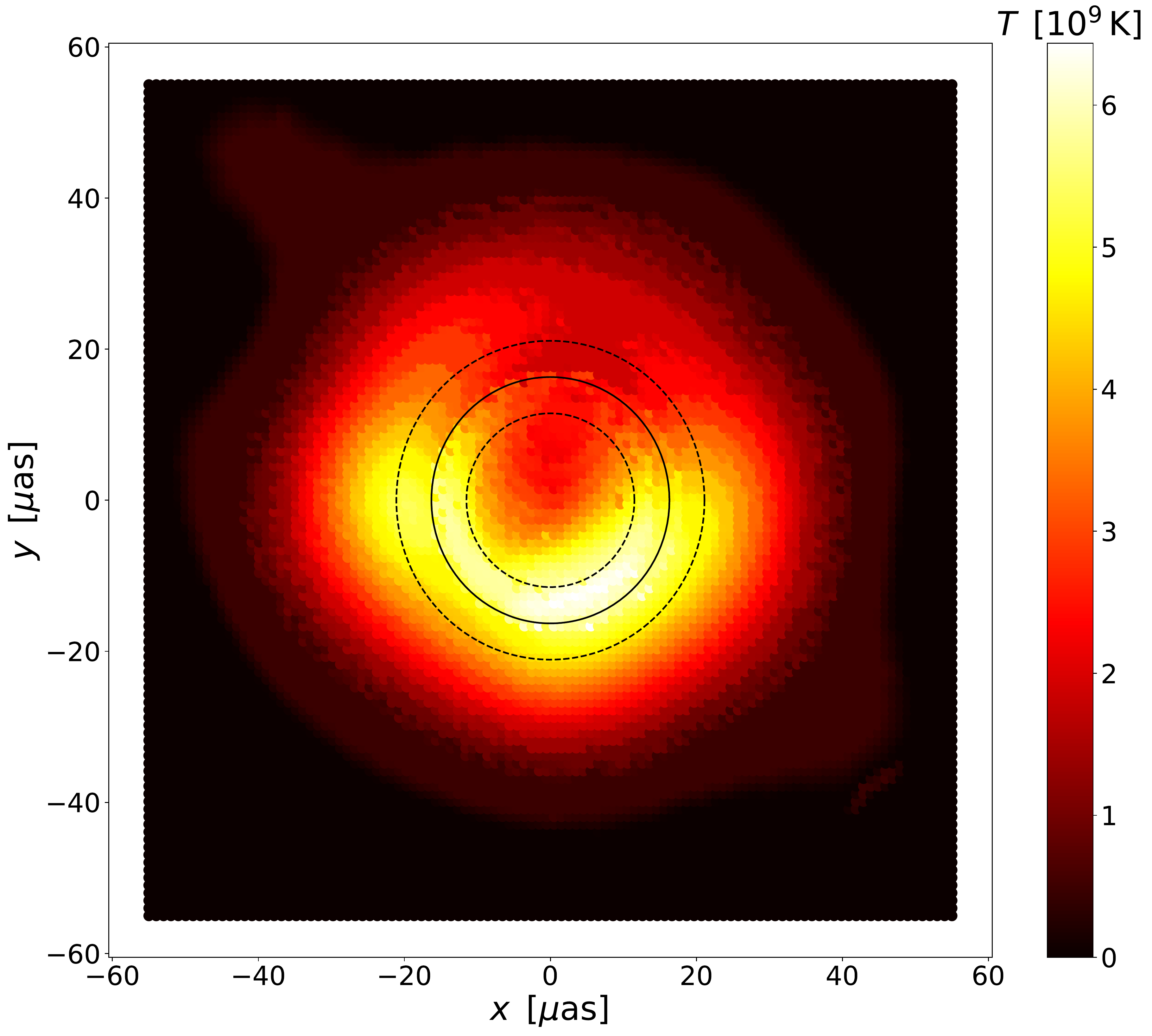}
\includegraphics[width=.49\textwidth]{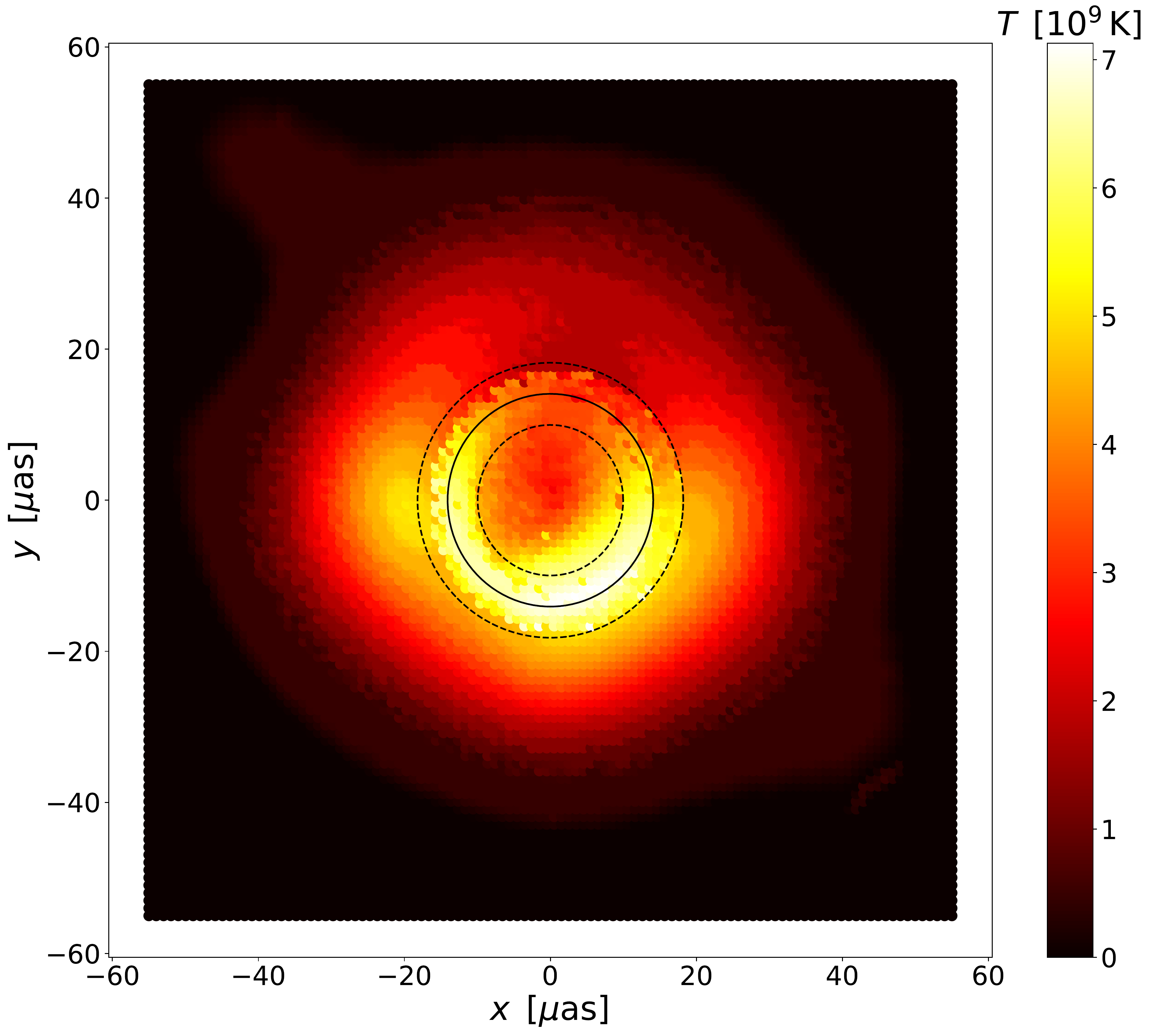}
\includegraphics[width=.49\textwidth]{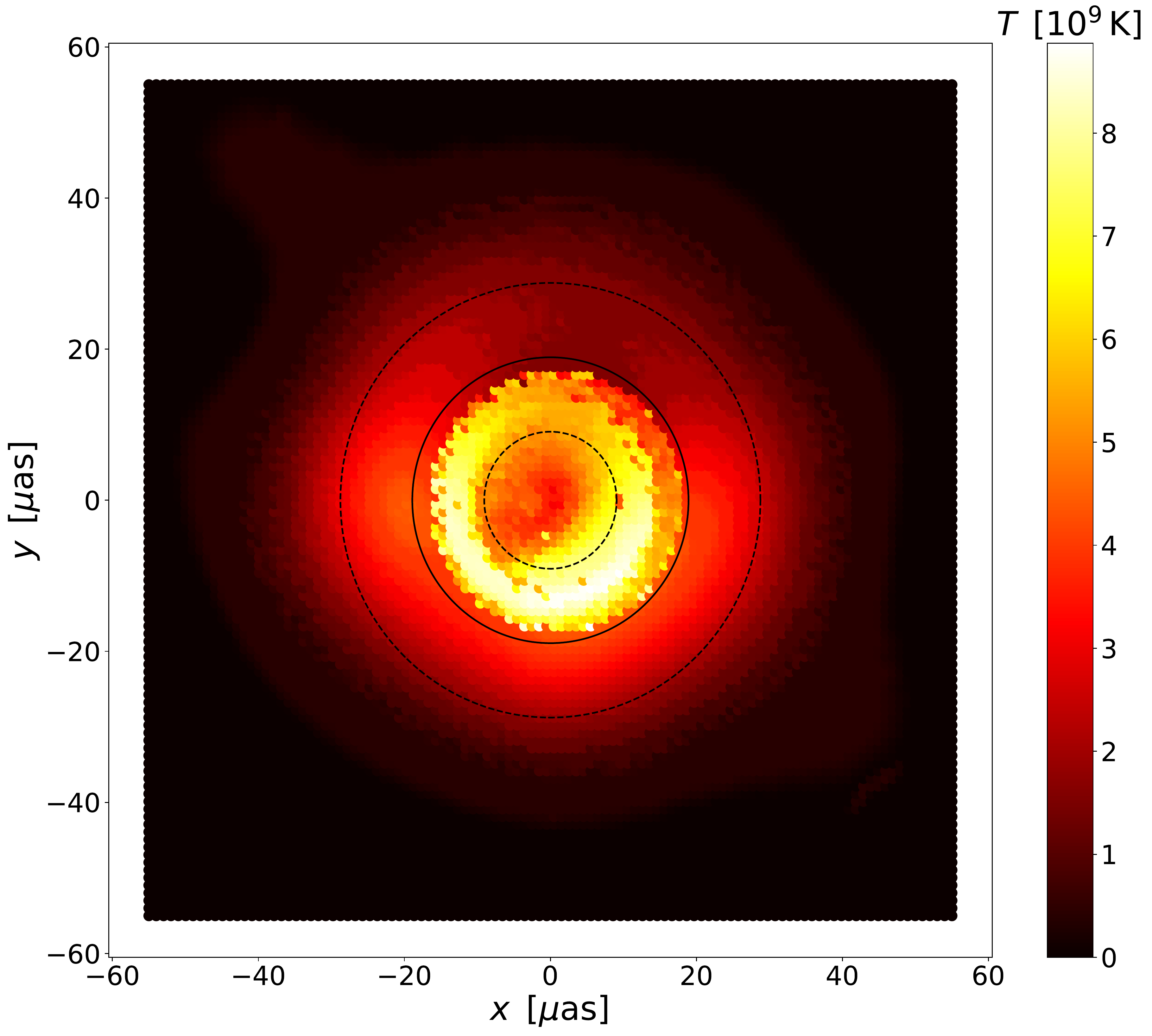}
\caption{\label{fig:EHT_added_l0} The added EHT image for $R_0 = 0.01, 0.2,
  0.4, 1$ (top left to bottom right) for $l=0$ and black hole spin $a=0.5$. The
  solid black line is half a ring diameter $d/2$ from the image center where
  $d$ is defined in eq.~\eqref{eq:d}. The dashed black lines are at radii
  $(d-\sigma_d)/2$ and $(d+\sigma_d)/2$, respectively where $\sigma_d$ is
  defined in~\eqref{eq:sigmad}.}
\end{figure}

To create the modified EHT images we implement the following procedure:
\begin{itemize}
  \item Ray trace the pixels $p \in P$ in the original EHT image back to a
    surface close to the horizon. This creates a "Close to the Horizon" (CTH)
    map at radius $r_{\rm CTH}$ that assigns the intensity of
    each pixel $I_p$ of the EHT image to a point on this surface $(r_{\rm CTH},
    \theta, \phi)$:
    \begin{equation}
      \Omega_p \equiv (\theta, \phi)_p = \Omega_p (I_p)\,.
    \end{equation}
    We choose $r_{\rm CTH}$ to be the photon ring radius, i.e.~the radius at
    which photons travel in unstable orbits. Hence, the horizon, reflection
    surface and CTH radii obey the hierarchy
    \begin{equation}
      r_\ssH < r_\ssR < r_{\rm CTH}\,.
    \end{equation}
    The CTH map allows us to assign intensities to rays that have been reflected
    close to the horizon in a modified theory of gravity as it gives us an idea
    where the rays we observe at the EHT originate in close proximity to the
    supermassive black hole.
  \item We now ray-trace the pixels turning on a reflection coefficient:
    Starting at the camera within the photon ring a ray would have simply
    fallen into the horizon in GR. However, with a
    non-vanishing reflection coefficient the ray gets (partially) reflected
    where it hits the surface $r_\ssR$ and ultimately escapes to infinity. What
    intensity do we assign to this particular ray? Somewhere between reflection
    and the escape to infinity, at $\Omega_{\rm CTH}$ the ray crosses the CTH
    surface. As the CTH map informs us about the origin of a rays intensity we
    can use this map to assign an intensity of $I_R$ to the reflected ray as
    the intensity $I_p$ of the closest point to crossing on the CTH surface:
    \begin{equation}
      I_R = R(\Omega_{\rm CTH}) \cdot \arg\left[ \min\left\{\left| \Omega_{\rm
      CTH} - \Omega_p(I_p)\right| \right\}_{p \in P}\right]\,,
    \end{equation}
    where the reflection coefficient $R$ is defined in eq.~\eqref{eq:Rl_def}.
  \item We can now add the two images: the unreflected original EHT image and 
    the reflected image created in the previous step. The image outside the 
    photon ring is not modified as those rays do not cross the reflection
    surface just outside the black hole horizon. The image is normalized such 
    that the sum of all pixel intensities is the same as in the original EHT
    image. Examples of these images can be found in
    Figure~\ref{fig:EHT_added_l0} for $l=0$ reflection and in
    Appendix~\ref{appendix:imagesl12} for higher multipole reflection
    coefficients $l=1,2$.
  \item Finally we run the image diagnostics discussed in
    Section~\ref{sec:EHTbackground} to determine if an image is consistent with
    the diagnostics of the observed image.
\end{itemize}
Note that this procedure also comes with a few caveats:
\begin{itemize}
  \item It is a perturbative approach in the sense that we can only constrain
	  small changes to the image as we infer the intensities of the
		reflected rays from the original (unperturbed) image. However,
		this assumption is justified as the EHT image is consistent
		with a Kerr black hole in GR.
  \item To trace the rays through the accretion disk in the vicinity of the
	  black hole, the medium has to be optically thin. This assumption is
		consistent with what is known about the accretion disk of
		M87~\cite{Narayan:1994xi, Narayan:1994is, Reynolds:1996is}.
  \item Finally, we assume the reference surface radiates with
	  equal intensity in all directions, {\it i.e.}~the intensity of the
		reflected rays only depends on where in the reference surface it
		starts (and not on the direction in which it travels).
\end{itemize}

A clear advantage of this approach is that we do not have to make additional
assumptions or make simulations (on top of those the EHT collaboration is
relying on already) about the environment close to the horizon of M87. That
being said the ultimate way to constrain these theories of modified gravity 
described by a reflection coefficient would be to incorporate reflection
into general relativistic magnetohydrodynamic (GRMHD) simulations such as those
relied upon by the EHT collaboration~\cite{Akiyama:2019fyp}.

\section{Results}
\label{sec:results}

\begin{figure}[!htb]
\centering
\includegraphics[width=.95\textwidth]{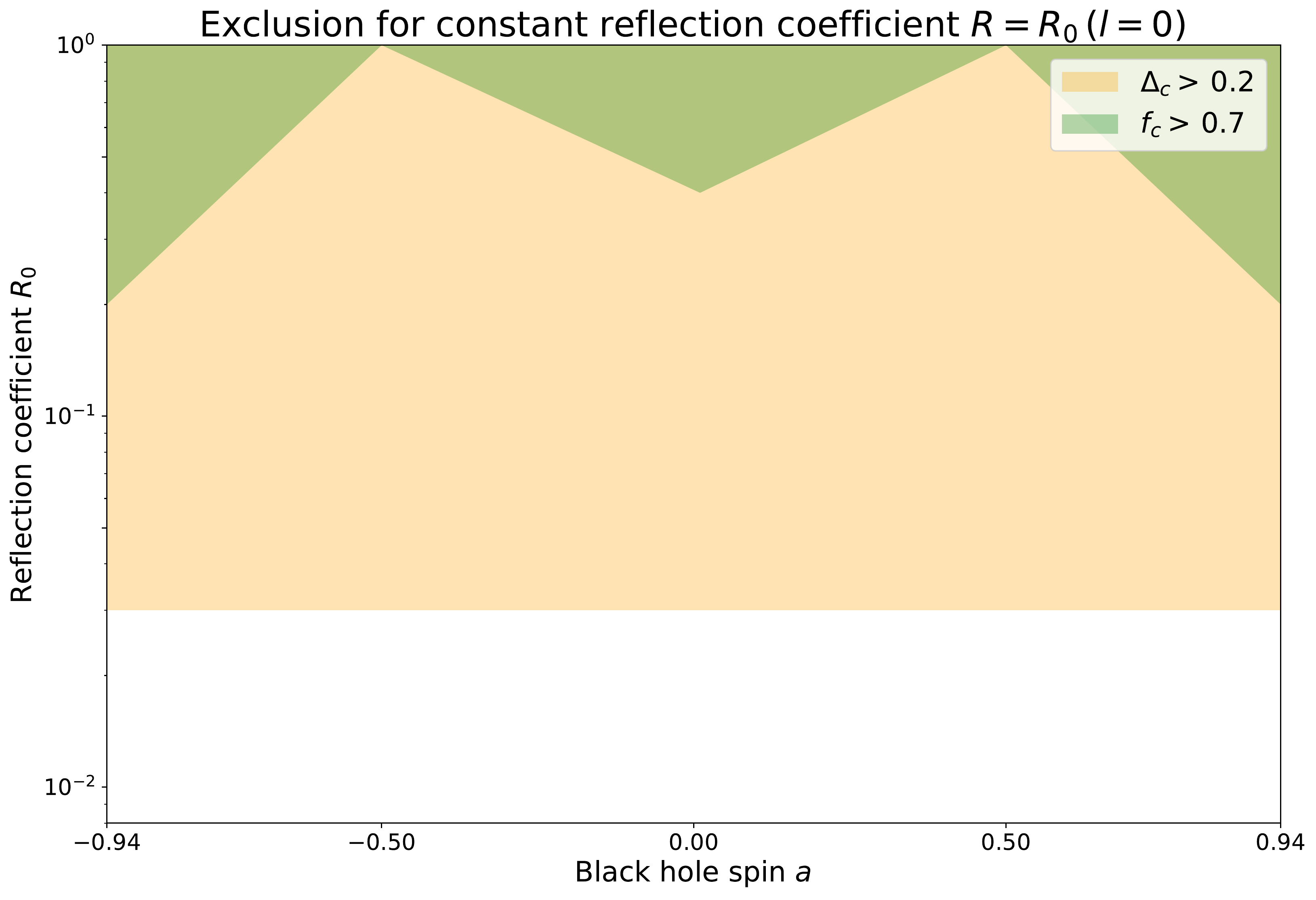}
\caption{\label{fig:exclusions_l0} Exclusion plot for constant reflection
        coefficient $l=0$. The coloured regions are excluded by
	$\Delta_c > 0.2$ and $f_c > 0.7$, respectively.$^{\ref{cl_footnote}}$}
\end{figure}

\begin{figure}[!htb]
\centering
\includegraphics[width=.95\textwidth]{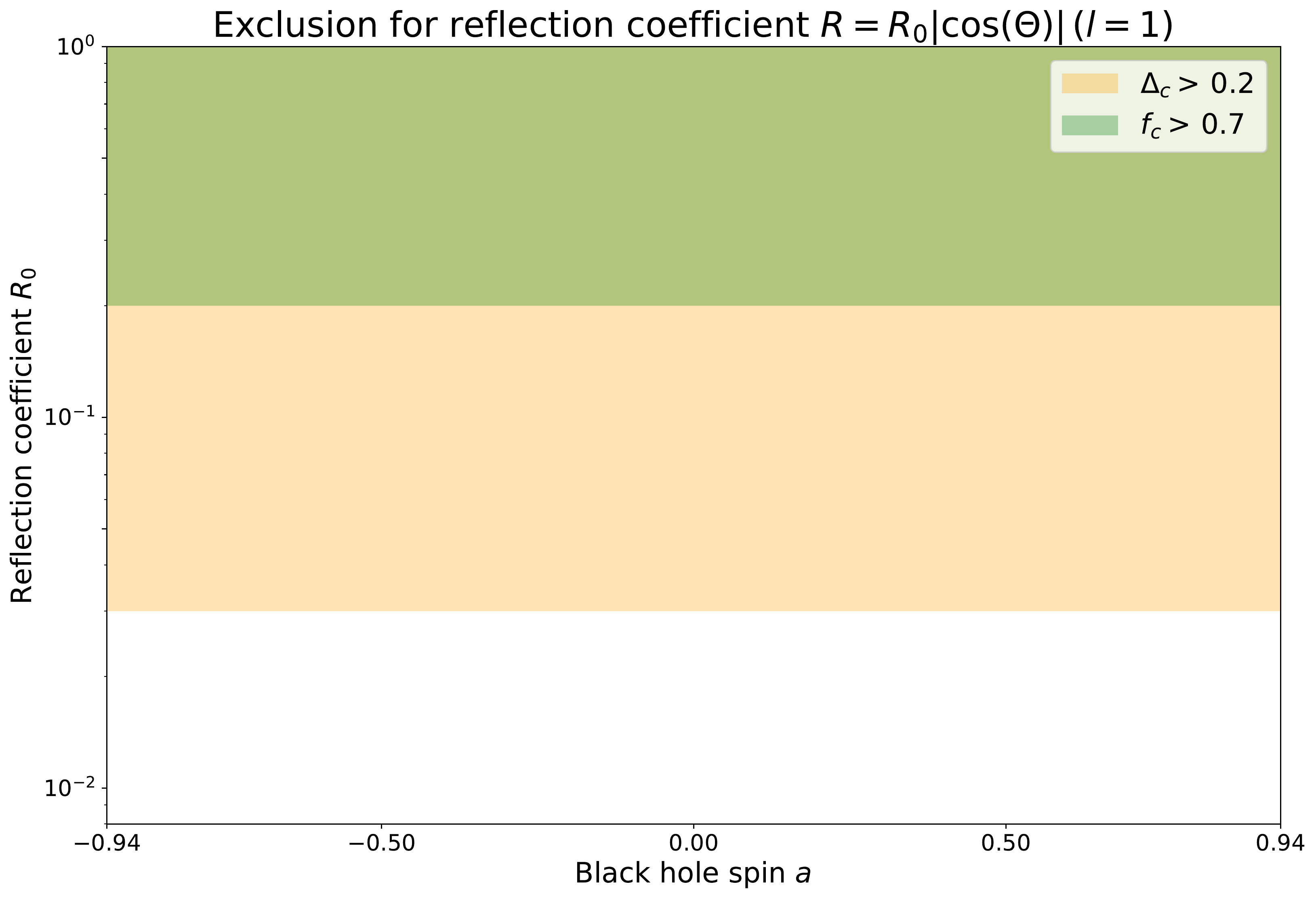}
\caption{\label{fig:exclusions_l1} Exclusion plot for the $l=1$ reflection
        coefficient. The coloured regions are excluded by
        $\Delta_c > 0.2$ and $f_c > 0.7$, respectively.$^{\ref{cl_footnote}}$}
\end{figure}

\begin{figure}[!htb]
\centering
\includegraphics[width=.95\textwidth]{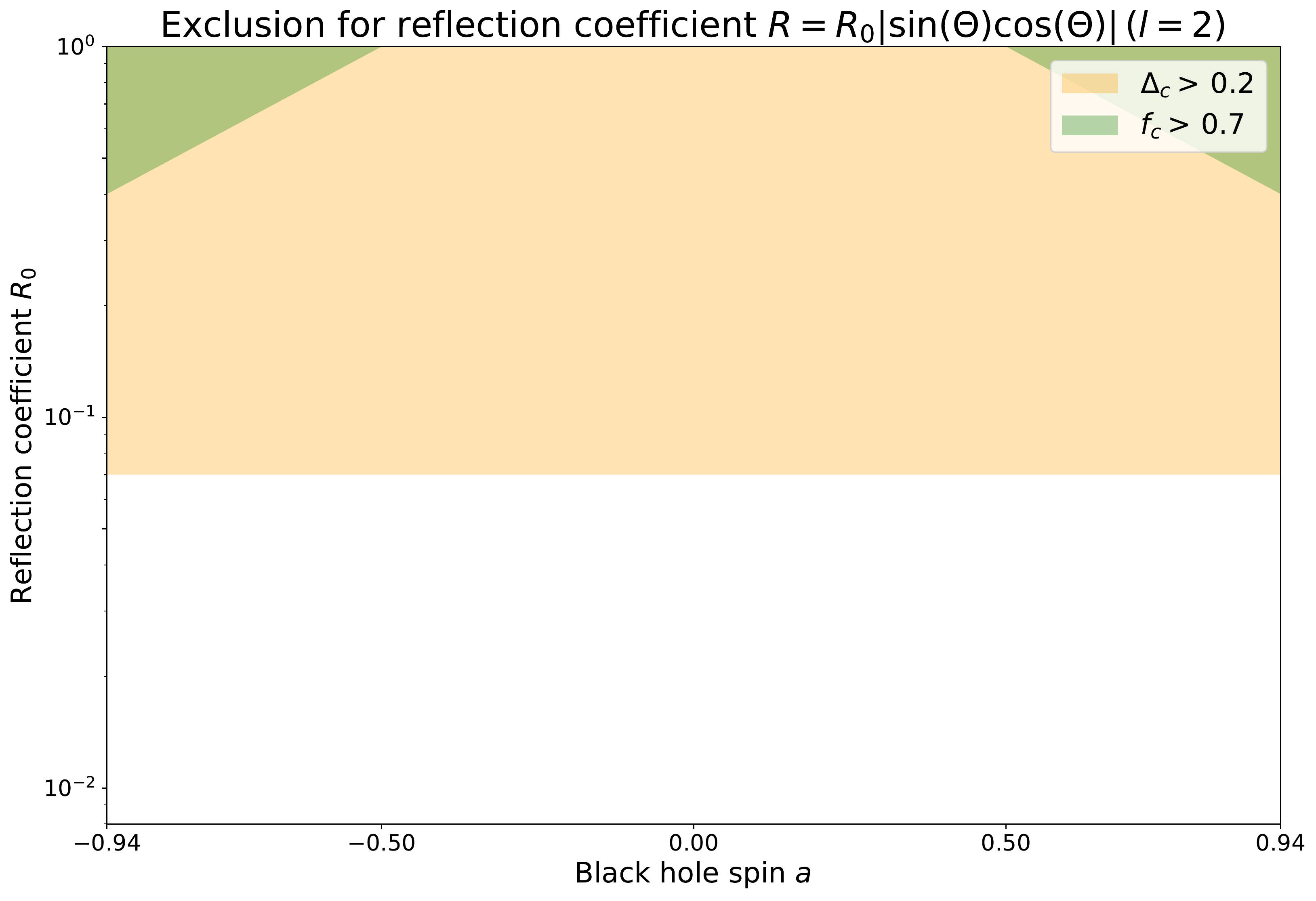}
\caption{\label{fig:exclusions_l2} Exclusion plot for the $l=2$ reflection
        coefficient. The coloured regions are excluded by
        $\Delta_c > 0.2$ and $f_c > 0.7$, respectively.$^{\ref{cl_footnote}}$}
\end{figure}

We scan over a set of models defined by the possible combinations of the
following parameters:
\begin{align}
	\begin{aligned}
		l &\in \left[0, 1, 2 \right]\quad {\rm and}\quad
		R_0 \in \left[10^{-3}, 5\cdot 10^{-3}, 10^{-2}, 3\cdot 10^{-2},
		7\cdot10^{-2}, 10^{-1}, 2\cdot 10^{-1}, 4\cdot10^{-1}, 
		1 \right]\,.
	\end{aligned}
\end{align}
One choice of $l$ and $R_0$ completely defines an effective reflection model
according to eq.~\eqref{eq:Rl_def}. Since the spin of M87 is to date not
determined~\footnote{The exception is that all models with $a=0$ considered
in~\cite{Akiyama:2019fyp} are excluded by too small jet power.} we also scan
over a variety of spin parameters, similar to those considered in
\cite{Akiyama:2019fyp}:
\begin{equation}
	a\in \left[-0.94, -0.5, 0.01, 0.5, 0.94\right]\,.
\end{equation}

In order to determine if a model is excluded, we evaluate the exclusion
criterion formulated at the end of Section~\ref{sec:ImageAnalysis} for every
combination of $l$, $R_0$ and $a$, i.e.~we test if the image characteristics
$d$, $w$, $\eta$ and $A$ are in agreement with the values in the first row of
Table~\ref{tab:image_diagnostics} and the upper bounds
in~\eqref{eq:digitized_upper_bounds} are not violated.

We find that the deviation from circularity $\Delta_c$ is the most constraining
property, as images with a sizable reflection coefficient tend to increase the
brightness towards the center of the image (see
e.g.~Figure~\ref{fig:EHT_added_l0}) which tends to increase $\sigma_d$ and
hence $\Delta_c$.\footnote{For the confidence level of exclusions from
$\Delta_c$, see Section~7.4 of~\cite{Akiyama:2019eap}. \cite{Akiyama:2019eap}
does not quote a precise confidence level of their upper bound $\Delta_c < 0.1$
(we use $\Delta_c < 0.2$ as out image quality is worse) but their Figure~18
suggests a confidence level $\gtrsim 90\%$.\label{cl_footnote}} Another effect
of increasing the brightness towards the image center is an increased
fractional central brightness $f_c$. As this quantity is in general less
constrained by the EHT observations, the exclusion bounds we find from $f_c$
are about an order of magnitude weaker than those from $\Delta_c$.  $d$, $w$,
$\eta$ and $A$ are 1-$\sigma$ compatible with the values in
Table~\ref{tab:image_diagnostics}, except some parameter points at $R_0=1$ that
are anyways excluded by $\Delta_c$ and $f_c$. Note that the errors we determine
for these quantities in Section~\ref{sec:ImageAnalysis} are rather large as we
are using a digitized image of the EHT observations.  Using the about three
times smaller errors on these quantities and stronger bounds on $\Delta_c$ and
$f_c$ from the original EHT image eq.~\eqref{eq:EHT_upper_bounds}, would lead
to more competitive bounds than from our digitized image.

Our bounds are summarized in Figures~\ref{fig:exclusions_l0} ($l=0$),
\ref{fig:exclusions_l1} ($l=1$), and \ref{fig:exclusions_l2} ($l=2$). For the
lowest multipole $l=0$, i.e.~angular independent reflection coefficient and the
first multipole $l=1$ which induces a $\cos \theta$ dependence on the polar
angle we find the bound
\begin{equation}
	R_0 \lesssim 3\cdot 10^{-2} \qquad {\rm for}\qquad l=0,1\,,
\end{equation}
for all possible spins $a$ we consider. For $l=0$, $f_c$ excludes high spin
models $|a|=0.94$ at $R_0 \lesssim 2\cdot 10^{-1}$ while for $l=1$, $R_0
\lesssim 2\cdot 10^{-1}$ for all values of $a$. 

For $l=2$, the constraint from $\Delta_c$ is weaker at
\begin{equation}
	R_0 \lesssim 7\cdot 10^{-2} \qquad {\rm for}\qquad l=2\,,
\end{equation}
while $f_c$ can only exclude models with $|a| = 0.94$ at $R_0 \lesssim 4\cdot
10^{-1}$. Via \eqref{lR_ordermag_eq} these constraints on $R_0$,
imply a bound $\ell \lesssim (7-70) \mu m$.

Note, that an interesting effect of significant reflection is
to focus more brightness inside the photon ring, i.e.~into the center of the
image. If reflection would be integrated into GRMHD simulations (as opposed to
our perturbative treatment) this might lead to smaller ring diameters for a
given mass of the supermassive black hole in M87. Since there is already a
tension with kinematic measurements preferring a lower
mass~\cite{Walsh:2013uua}, this could lead to even more stringent constraints
on modified gravity theories when these kinematic measurements are taken into
account.

\section{Conclusions}
\label{sec:conclusions}

In this work, we constrained modifications of GR using the EHT
observations of M87~\cite{Akiyama:2019cqa}. Many interesting extensions of
GR~\cite{Almheiri:2012rt, Susskind:2012uw, Lunin:2001jy} can be
effectively described by a reflection coefficient that specifies how much of an
ingoing wave is reflected into an outgoing wave at a surface close to where the
event horizon of a classical black hole would be. The phenomenology of such a
reflecting black hole system can be coherently described in a PPEFT framework
as discussed in~\cite{Burgess:2018pmm}.

If the outgoing/reflected wave reaches the detector, in this case the EHT, it
modifies the image. We constructed images from modified gravity theories using
a ray tracing algorithm and the EHT observations to assign intensities to the
reflected rays. While our approach is useful to constrain modified gravity
theories that only effect the EHT observations perturbatively - which is
justified since the observations are consistent with a Kerr black hole - it
would be a natural next step to include reflection into GRMHD simulations that
are used by the EHT~\cite{Akiyama:2019fyp}.

We used image diagnostics such as the image rings' deviation from circularity
and fractional central brightness to determine if the modified images from
reflection are consistent with the EHT observations. We find that we can
constrain the RG invariant reflection coefficient $R$ to be less than
$1\,-\,10\%$ and $\ell < (7-70) \mu m$ with a weak dependence on black hole spin and angular dependence
of $R$. We expect these constrains to improve with the original EHT dataset and
GRMHD simulations.

\acknowledgments

We thank Luis Lehner, Peter Hayman, Greg Kaplanek and Laszlo Zalavari for
helpful discussions. This work was partially supported by funds from the
Natural Sciences and Engineering Research Council (NSERC) of Canada. Research
at the Perimeter Institute is supported in part by the Government of Canada
through NSERC and by the Province of Ontario through MRI. 

\appendix

\section{Reflection coefficients and EFT couplings}
\label{appendix:Rh}

In this section, we briefly summarize how the RG invariant reflection
coefficient $R$ is related to the EFT coupling $h_1$ of eq.~\eqref{horaction}
and the new-physics RG-invariant length-scale $\ell$. For a more in-depth
discussion see~\cite{Burgess:2018pmm} and \cite{Plestid:2018qbf}.

Exotic UV physics that is localized near the horizon affects physics far from
the horizon through the changes it makes to the near-horizon boundary condition
experienced by any external low-energy `bulk' fields used to probe the
near-horizon regime. In the EFT formalism of refs.~\cite{Burgess:2016lal,
Burgess:2016ddi, Burgess:2017ekj}, any particular type of modified near-horizon
boundary condition is captured in terms of a surface contribution, 
\be
   S_b = \int_\Sigma \exd^3x \; \mathfrak{L}_b \,, 
\ee
to the low-energy effective action whose presence `tells' the low-energy theory
that the boundary condition gets modified.\footnote{This is similar in spirit
to the black-hole `membrane paradigm' \cite{Thorne:1986iy} used in black-hole
astrophysics. One way of thinking about the EFT of \cite{Burgess:2018pmm} is as
a theoretical framework in terms of which this paradigm can be derived
(including systematic corrections) and adapted to extensions of GR.} The
surface $\Sigma$ can be anywhere, and the effective couplings in $S_b$ depend
on $\Sigma$ in precisely the way they must to ensure that observables are
$\Sigma$-independent. For UV physics localized near the would-be horizon the
cleanest split between UV and other effects arises if $\Sigma$ is chosen in the
near-horizon regime (but outside the region where the UV physics is important).
In this language any freedom of choice in the nature of $R$ appears as the
freedom to choose effective couplings within the boundary action $S_b$.  

When the dust settles, physical quantities (like reflection coefficients) are
RG invariants in the sense that they depend on effective couplings in a way
that is $\Sigma$-independent. This implies they can be expressed in terms of
RG-invariant characterizations of coupling-constant flow. For the simplest
couplings --- like that of \pref{horaction} --- RG evolution turns out to be
labeled by two parameters: an RG-invariant phase, $e^{i\Theta_\star}$, and
length scale, $\ell$, via a process similar to dimensional transmutation.
$\ell$ and $\Theta_\star$ are RG-invariant parameters that uniquely specify a
particular RG flow line, and physically parameterize the relative amplitudes of
reflection and absorption as well as any phase shift associated with
reflection. Of these only $\ell$ is needed to determine $|R|$, so this is all
that is required in the discussion of the main text.

The reflection coefficient generally not only depends on the UV
physics but grey-body factors and normalization constants of the full mode
functions (which don't depend on $h_1$, $\epsilon$, {\it etc.}). These
grey-body factors are crucial when the energy of the scattered wave is similar
to the size of the potential barrier, which is determined by the parts of the
black hole gravitational potential that are not the inverse-square potential
that dominates near the horizon. For the EHT however, the scattered wave has
a much higher energy than the gravitational potential. The EHT wavelength of
1.3 mm is much smaller than the horizon scale which is $\mathcal{O}(10^{13})$
m. Hence, the wave does effectively not 'see' the gravitational potential
except close to the horizon where the inverse-square potential dominates.

As a result of this hierarchy of scales $\omega \gg 1/r_s$, the physical
reflection coefficient $R$ discussed in this work is to very good approximation
given by the inverse-square reflection coefficient of the EFT, defined
in~\cite{Burgess:2018pmm} as
\begin{equation}
  \label{eq:Rinvsquare}
  R \simeq |R_{\rm inv-sq}| = \left| \frac{\zeta - \lambda(\epsilon)}{\zeta + 
   \lambda(\epsilon)} \left( 2 i \tilde k \epsilon \right)^{\zeta}\right|\,,
\end{equation}
where, for boundary conditions in Kerr spacetime specified at the radial
coordinate $r = r_\ssH + \epsilon$ (in Boyer-Lindquist coordinates -- see
\pref{BoyerLind}), 
\begin{equation}
    \lambda(\epsilon) = \frac{1}{2\pi\epsilon}\left[ \frac{\rho(r, \theta) 
  h_1(\theta, \epsilon)}{\sqrt{\Delta(r)}}\right]_{r = r_\ssH + \epsilon} + 1\,,
\end{equation}
where $\theta$-independent $R$ requires a $\theta$-dependence of $h_1(\theta,
\epsilon)$ such that $ \lambda(\epsilon)$ is independent of $\theta$. The
constant $\tilde k$ is given in (2.28) in~\cite{Burgess:2018pmm} and in the limit
$\omega r_s \to \infty$ becomes
\begin{equation}
  \tilde k \simeq \omega\,. 
\end{equation}
For a
Kerr black hole with angular momentum parameter $a$ and mass $M$, the constant
$\zeta$ is given as
\begin{equation}
  \zeta = 2\left\{\frac{s^2}{4} - \frac{(\omega r_s r_\ssH - am)^2 + i s \left[ 
  m a \sqrt{r_s^2-4a^2} - (r_\ssH^2 - a^2)r_s \omega\right]}{r_s^2-4a^2} 
  \right\}^{1/2}\,,
\end{equation}
where $r_s = 2GM$. This depends both on the spin $s$  of the bulk field (where
$s=0$ is a Klein-Gordon scalar, $s=1$ an electromagnetic field and $s=2$ a
spin-two metric fluctuation), the solutions' energy $\omega$ and magnetic
quantum number $m$. In the Schwarzschild ($a \to 0$) limit this becomes 
\begin{equation}
  \zeta \to  s + 2i \omega\, r_s  \qquad \hbox{(Schwarzschild limit)}\,.
\end{equation}

Since $R$ is an RG invariant, the dependence of $h_1$ on $\epsilon$ is
precisely such that it cancels the $\epsilon$ dependence of the remaining terms
on the RHS of eq.~\eqref{eq:Rinvsquare}. Alternatively, this can be described by
relating $R$ to the RG invariant length scale $\ell$, defined {\it
e.g.}~as the scale for which $  \lambda(\epsilon = \ell) = 0$. For the
phenomenology explored in this work, only $|R_{\rm inv-sq}|$ is relevant, and is
related to $\ell$ via
\begin{equation}
  \label{Rofzeta}
  R \simeq |R_{\rm inv-sq}| = \left( 2 \omega \ell \right)^{{\rm Re}(\zeta)}\,.
\end{equation}
Hence, $\ell \rightarrow 0$ leads to $R \rightarrow 0$,
i.e.~the vanilla GR limit. As discussed in~\cite{Burgess:2018pmm}, this is
the IR fixed point of the effective theory.

We can use \eqref{Rofzeta} to obtain a bound on $\ell$:
\begin{equation}
 \label{lR_ordermag_eq}
  R \simeq \frac{\ell}{655 \mu {\rm m}} \,,
\end{equation}
using $Re(\zeta) = s = 1$ for photons and using the EHT observation wavelength
of 1.3 mm.  Eq.~\eqref{lR_ordermag_eq} can be directly used to obtain an order
of magnitude constraint on $\ell$ from a bound on $R$ as described in
Section~\ref{sec:results}.

\section{Kerr geodesics}
\label{appendix:Kerrgeodesic}

Here, we list the expressions $F^i(x^j, p^j)$ necessary to solve the Kerr
geodesic differential equation~\eqref{eq:geodesicwesolve}:
\begin{align}
\begin{aligned}
F^r &= \frac{1}{2C} \left[\frac{\partial A}{\partial r}\dot t^2 - 
	2 \frac{\partial B}{\partial r}\dot t p^\phi -
	\frac{\partial C}{\partial r} (p^r)^2 -
        2 \frac{\partial C}{\partial \theta} p^\theta p^r +
	\frac{\partial D}{\partial r} (p^\theta)^2 +
	\frac{\partial F}{\partial r} (p^\phi)^2  \right]\,,\\
F^\theta &= \frac{1}{2D} \left[\frac{\partial A}{\partial \theta}\dot t^2 - 
	2 \frac{\partial B}{\partial \theta}\dot t p^\phi +
	\frac{\partial C}{\partial \theta} (p^r)^2 -
        2 \frac{\partial D}{\partial r} p^\theta p^r -
	\frac{\partial D}{\partial \theta} (p^\theta)^2 +
	\frac{\partial F}{\partial \theta} (p^\phi)^2  \right]\,,\\
F^\phi &= \frac{1}{G} \left[E \frac{\partial H}{\partial r} p^r + 
	E \frac{\partial H}{\partial \theta} p^\theta -
        \frac{\partial G}{\partial r} p^r p^\phi -
        \frac{\partial G}{\partial \theta} p^\theta p^\phi \right]\,,
\end{aligned}
\end{align}
where $\dot t$ is eliminated via energy conservation
\begin{equation}
	\dot t = \frac{E}{A} + H p^\phi\,,
\end{equation}
and
\begin{align}
\begin{aligned}
	A &= 1 - \frac{r_s r}{\rho^2}\,, \qquad 
	B = \frac{a \,r\, r_s \sin^2 \theta}{\rho^2}\,, \qquad
	C = \frac{\rho^2}{\Delta}\,, \qquad
	D = \rho^2\,,\\
	F &= \sin^2\theta \left(r^2 + a^2 + 
	  \frac{a \,r\, r_s \sin^2 \theta}{\rho^2} \right)\,,\qquad
        G = F - \frac{B^2}{A} 
	  = \sin^2\theta \left(r^2 + a^2 + 
	  \frac{a \,r\, r_s \sin^2 \theta}{\rho^2 - r\, r_s} \right)\,,\\
	H &= \frac{B}{A} = \frac{a \,r\, r_s \sin^2 \theta}{\rho^2 - r\, r_s}\,.
\end{aligned}
\end{align}

\section{Images for multipoles $l=1,2$}
\label{appendix:imagesl12}

\begin{figure}[!htb]
\centering
\includegraphics[width=.49\textwidth]{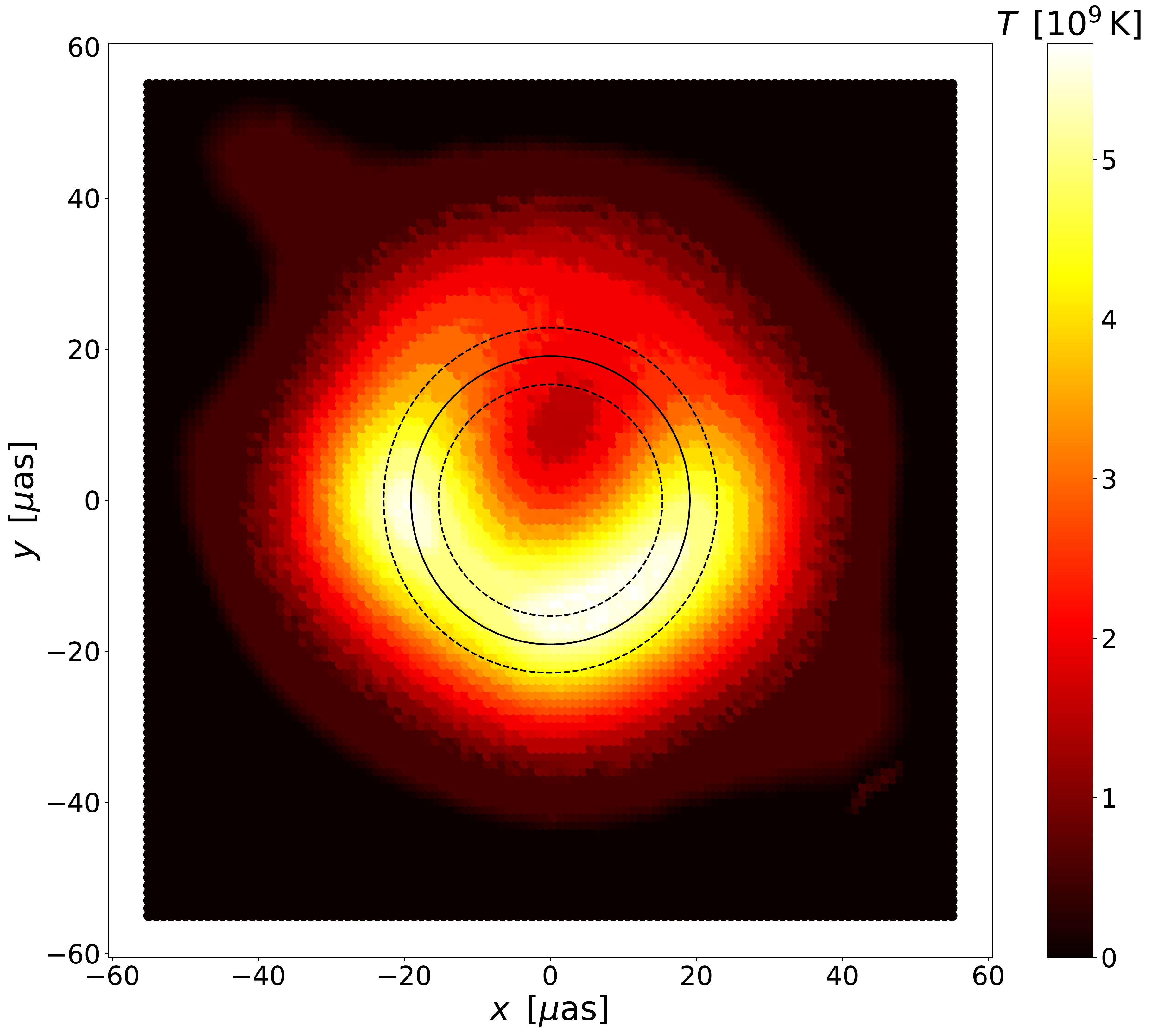}
\includegraphics[width=.49\textwidth]{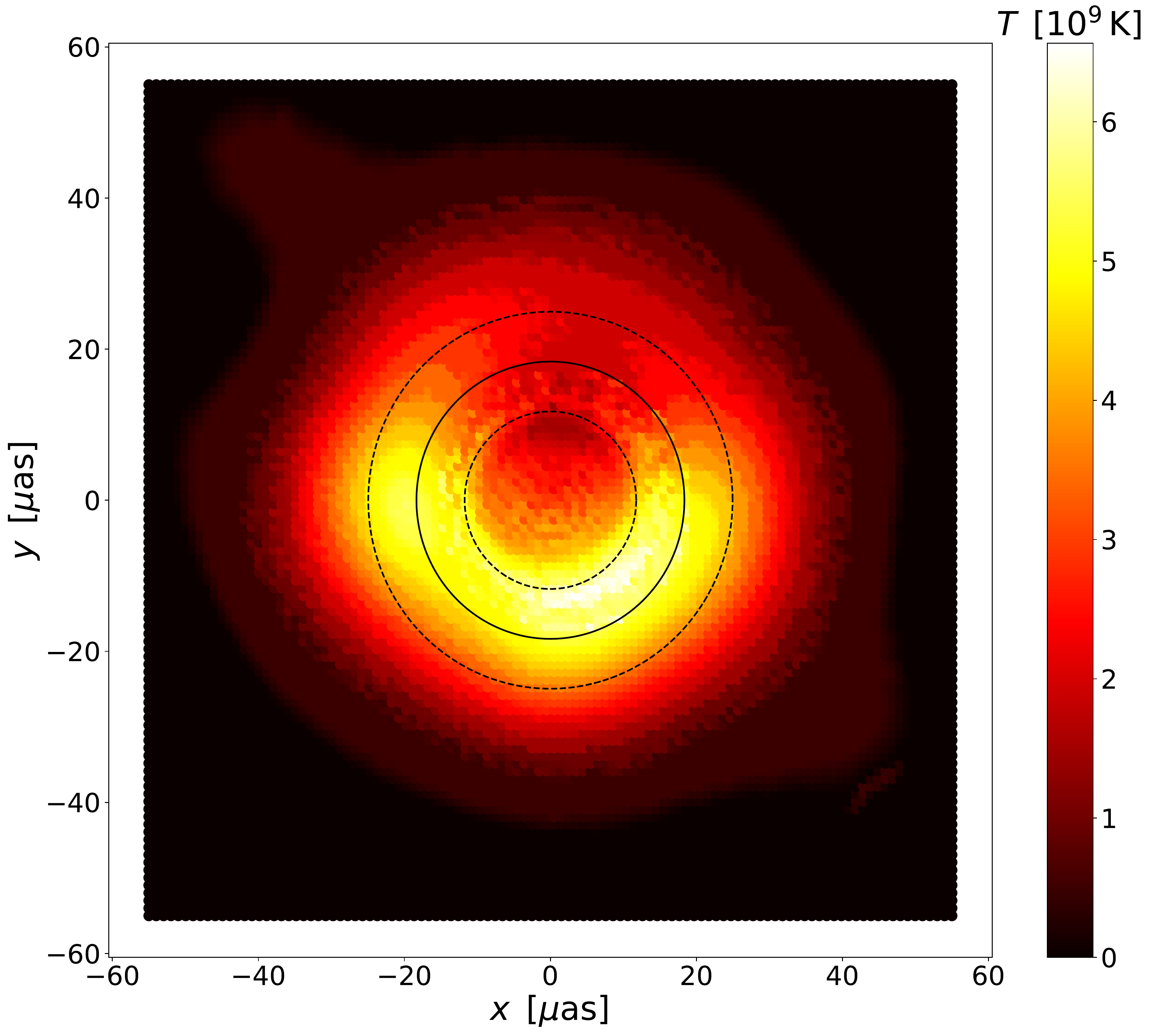}
\includegraphics[width=.49\textwidth]{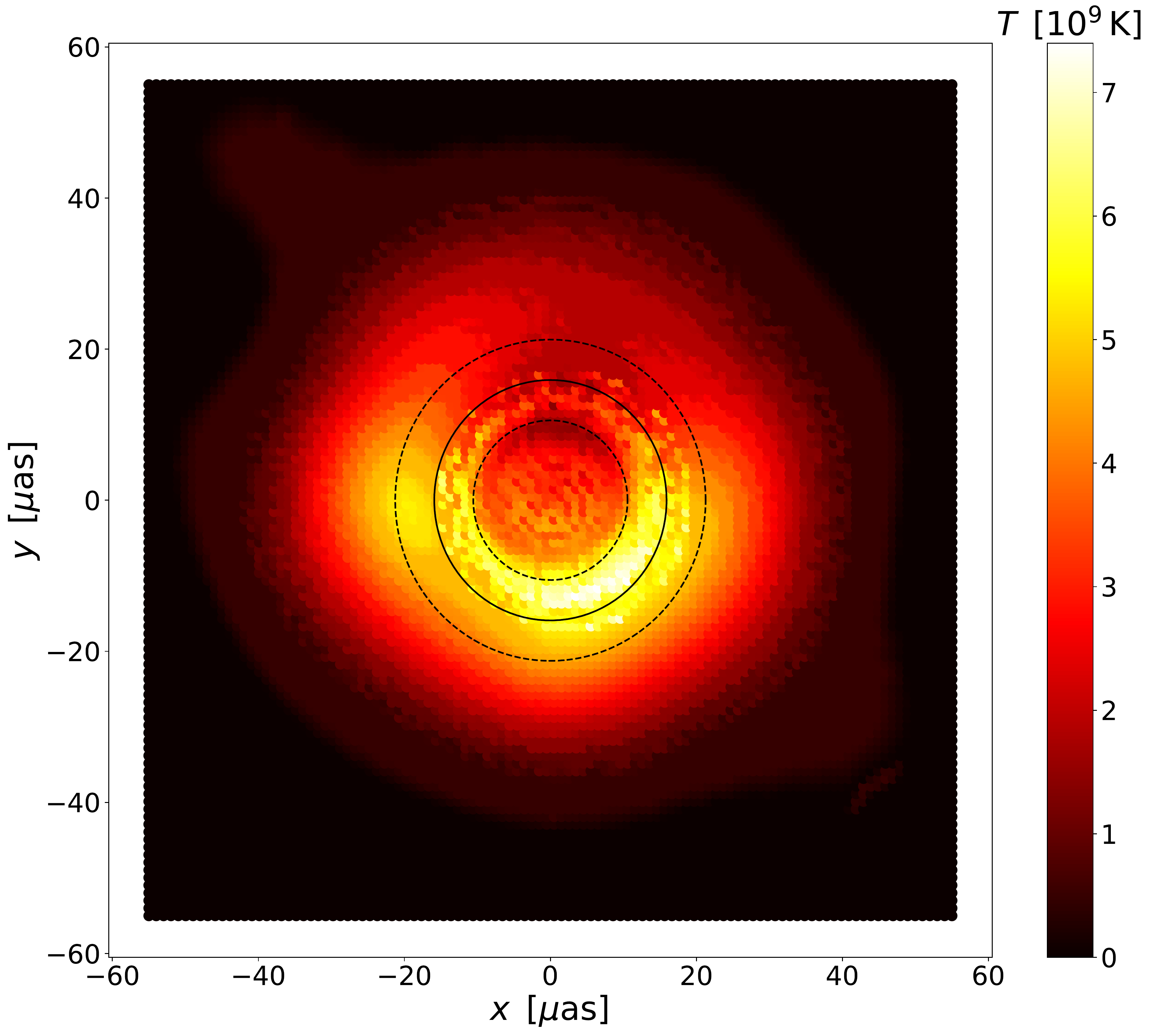}
\includegraphics[width=.49\textwidth]{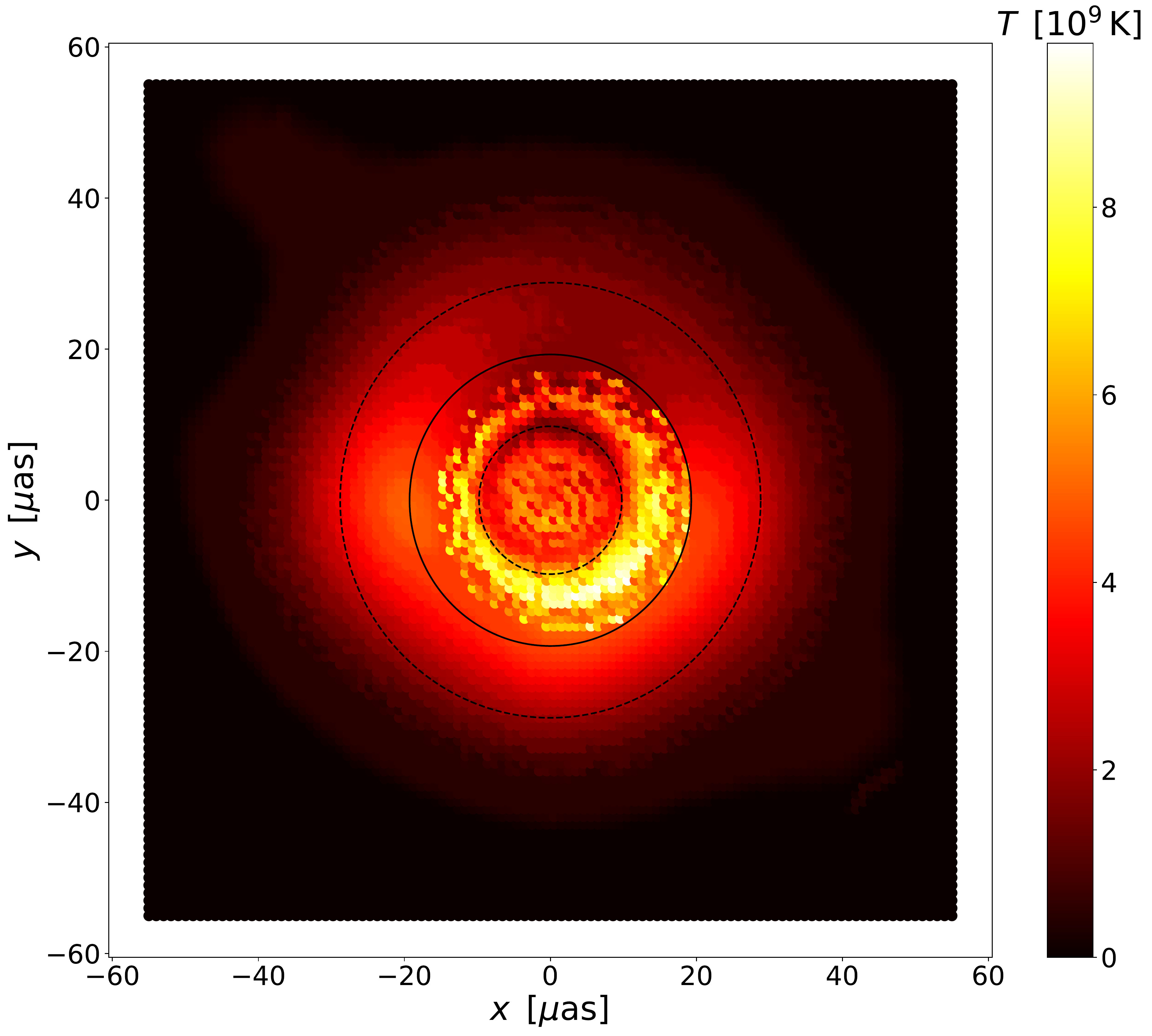}
\caption{\label{fig:EHT_added_l1} The added EHT image for $R_0 = 0.01, 0.2,
  0.4, 1$ (top left to bottom right) for $l=1$ and black hole spin $a=0.94$.
  The solid black line is half a ring diameter $d/2$ from the image center
  where $d$ is defined in eq.~\eqref{eq:d}. The dashed black lines are at radii
  $(d-\sigma_d)/2$ and $(d+\sigma_d)/2$, respectively where $\sigma_d$ is
  defined in~\eqref{eq:sigmad}.}
\end{figure}

\begin{figure}[!htb]
\centering
\includegraphics[width=.49\textwidth]{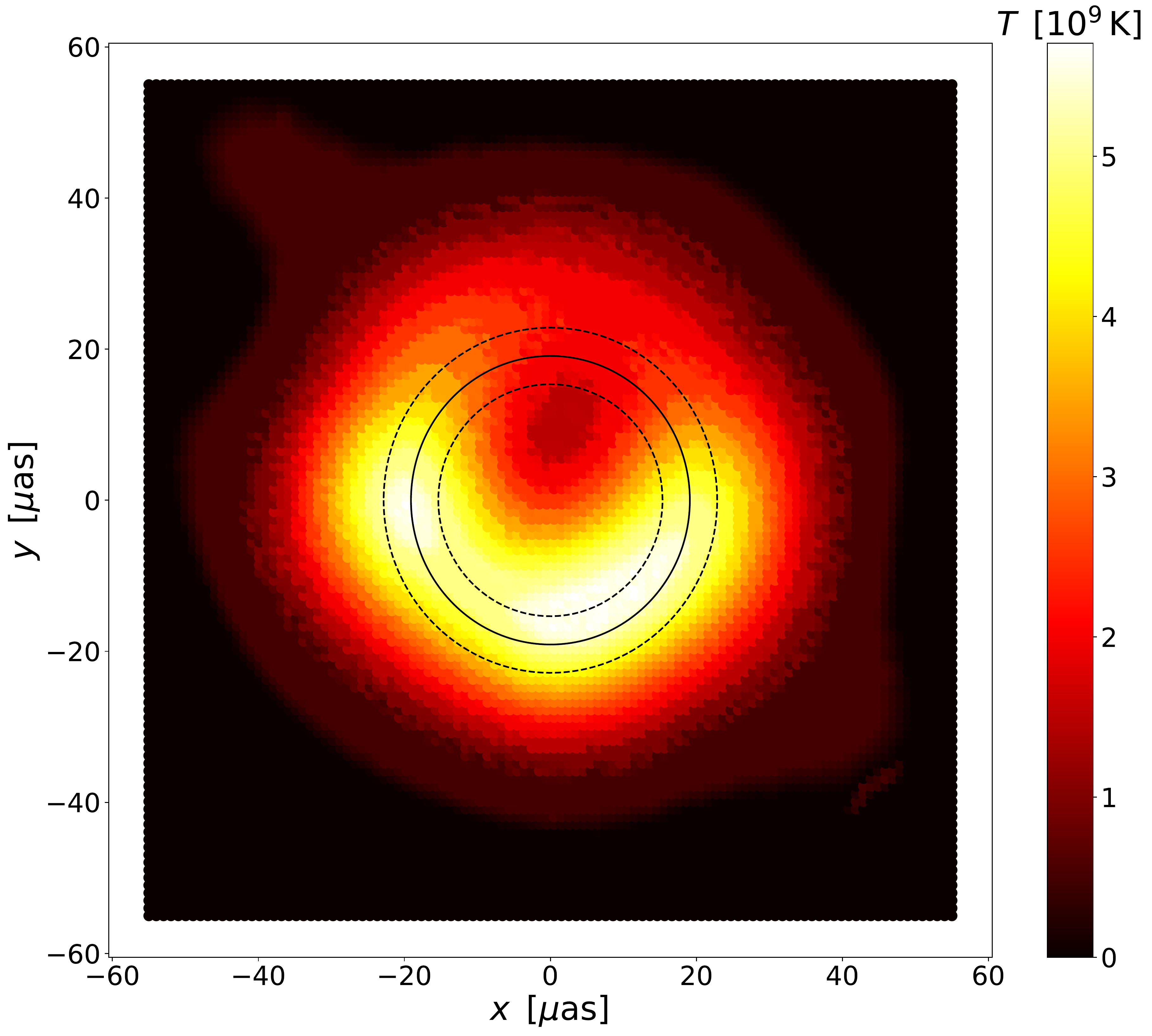}
\includegraphics[width=.49\textwidth]{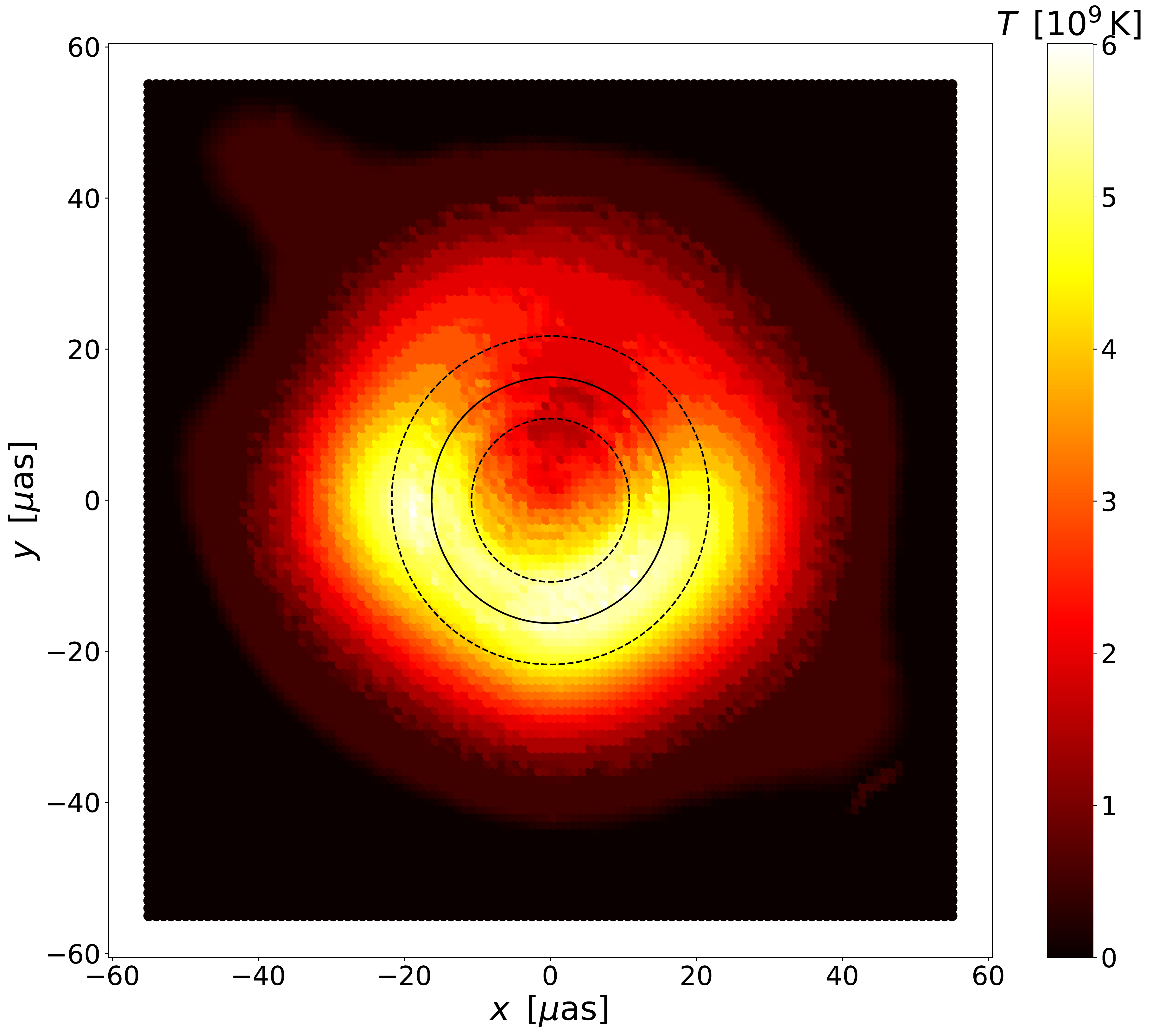}
\includegraphics[width=.49\textwidth]{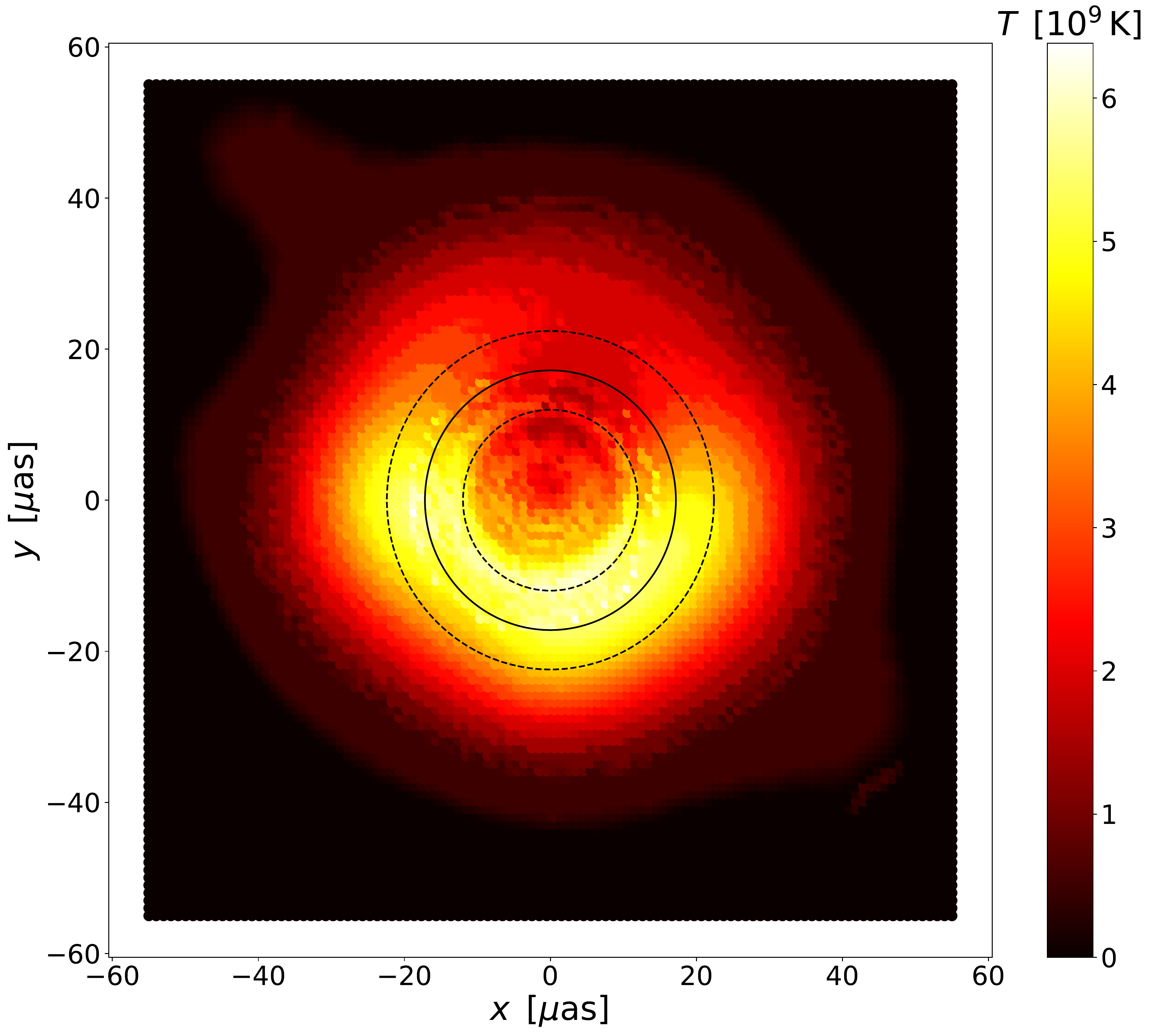}
\includegraphics[width=.49\textwidth]{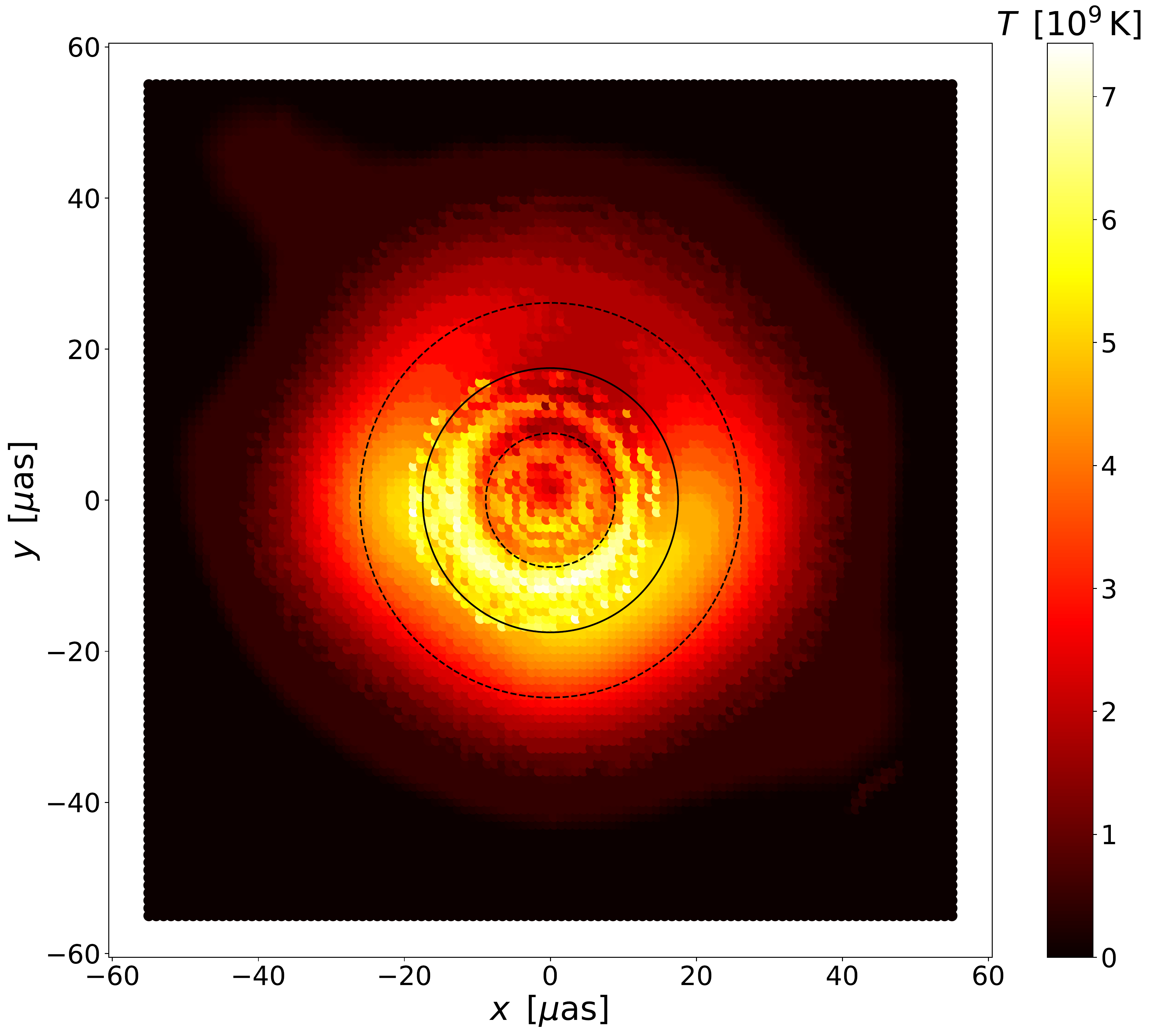}
\caption{\label{fig:EHT_added_l2} The added EHT image for $R_0 = 0.01, 0.2,
  0.4, 1$ (top left to bottom right) for $l=2$ and black hole spin $a=-0.94$.
  The solid black line is half a ring diameter $d/2$ from the image center
  where $d$ is defined in eq.~\eqref{eq:d}. The dashed black lines are at radii
  $(d-\sigma_d)/2$ and $(d+\sigma_d)/2$, respectively where $\sigma_d$ is
  defined in~\eqref{eq:sigmad}.}
\end{figure}

Here we show some examples of EHT images for higher multipole reflection
coefficients $l=1$ (Figure~\ref{fig:EHT_added_l1}) and $l=2$
(Figure~\ref{fig:EHT_added_l2}). The procedure how to create these images is
described in Section~\ref{sec:imagecreation}.

\clearpage
\bibliographystyle{JHEP.bst}
\bibliography{EHT_EFT_v4}
\end{document}